\pdfoutput=1
\documentclass[acmsmall]{acmart}

\usepackage{tikz}
\usepackage{subcaption}
\usepackage[linesnumbered, ruled, vlined]{algorithm2e}
\usepackage{xspace}
\usepackage{soul}
\usepackage{tabularx}
\usepackage{multirow}
\usepackage{array}
\usepackage{pifont}
\graphicspath{{figures/}}

\newcolumntype{Y}{>{\centering\arraybackslash}X}
\newcommand{\cmark}{\ding{51}}

\newcommand{\secref}[1]{Section~\ref{#1}}
\newcommand{\figref}[1]{Figure~\ref{#1}}
\newcommand{\tabref}[1]{Table~\ref{#1}}

\newcommand{\algoref}[1]{Algorithm~\ref{#1}}

\let\oldnl\nl
\newcommand{\nonl}{\renewcommand{\nl}{\let\nl\oldnl}}
\definecolor{comment-color}{rgb}{0.5,0.1,0.1}
\newcommand{\codeComment}[1]{\textnormal{\color{comment-color}{\textit{\#
#1}}}\unskip}
\sethlcolor{lightgray!75}
\SetAlFnt{\small}\SetAlCapFnt{\small}\SetAlCapNameFnt{\small}
\newcommand{\shl}[1]{\hl{\mbox{#1}}} %

\newcommand{\us}{\textmu s\xspace}
\newcommand{\proposed}{Nemo\xspace}
\newcommand{\twotree}{2-Tree\xspace}
\newcommand{\twotreeretain}{TwoTree-RI\xspace}
\newcommand{\twotreeevict}{TwoTree-EI\xspace}
\newcommand{\btree}{B+Tree\xspace}
\newcommand{\btreeretain}{B+Tree-RI\xspace}
\newcommand{\btreeevict}{B+Tree-EI\xspace}
\newcommand{\indexevict}{EI\xspace}
\newcommand{\indexretain}{RI\xspace}
\newcommand{\pcl}{PCL\xspace}

\newcommand{\boldtext}[1]{\noindent\textbf{#1}}

\DeclareRobustCommand{\circleb}[1]{%
  \tikz[baseline=(char.base)]{%
    \node[shape=circle, fill=black, inner sep=.5pt] (char)
         {\textcolor{white}{\small #1}};%
  }%
}

\begin{document}

\setcopyright{cc}
\setcctype{by}
\acmJournal{PACMMOD}
\acmYear{2026} \acmVolume{4} \acmNumber{6 (SIGMOD)} \acmArticle{362}
\acmMonth{12} \acmDOI{10.1145/3856357}

\title{Nemo: Efficient Caching with Index Pushdown for Disaggregated OLTP Databases}

\author{Wenjie Hu}
\orcid{0009-0004-7339-3594}
\affiliation{%
  \institution{University of Wisconsin-Madison}
  \city{Madison}
  \state{Wisconsin}
  \country{USA}
}
\email{wenjiehuhippo@cs.wisc.edu}

\author{Xinjing Zhou}
\orcid{0009-0008-5949-6553}
\affiliation{%
  \institution{Massachusetts Institute of Technology}
  \city{Cambridge}
  \state{Massachusetts}
  \country{USA}
}
\email{xinjing@mit.edu}

\author{Zhihan Guo}
\orcid{0009-0000-5729-698X}
\authornote{Work done while at the University of Wisconsin-Madison.}
\affiliation{%
  \institution{Salesforce, Inc.}
  \city{San Francisco}
  \state{California}
  \country{USA}
}
\email{zhihan.scarlet.guo@gmail.com}

\author{Xiangpeng Hao}
\orcid{0009-0001-9554-9084}
\affiliation{%
  \institution{University of Wisconsin-Madison}
  \city{Madison}
  \state{Wisconsin}
  \country{USA}
}
\email{xiangpeng.hao@wisc.edu}

\author{Xiangyao Yu}
\orcid{0009-0001-0785-2519}
\affiliation{%
  \institution{University of Wisconsin-Madison}
  \city{Madison}
  \state{Wisconsin}
  \country{USA}
}
\email{yxy@cs.wisc.edu}

\renewcommand{\shortauthors}{Wenjie Hu, Xinjing Zhou, Zhihan Guo, Xiangpeng Hao, \& Xiangyao Yu}

\begin{abstract}
Storage-disaggregated databases decouple compute from storage, improving elasticity but suffering from network bottlenecks. Most cloud databases use page-based B+trees and buffer pools. Under disaggregation, each index traversal incurs serial network round-trips. Page-granularity transfers and caching also amplify network traffic and lower cache efficiency. Pushing down the full index to storage and caching only hot records at compute is appealing, but an efficient design must (1) provide a holistic record cache that accelerates both point and range queries, (2) mitigate the negative search penalty of confirming non-existence via remote probes, and (3) enforce phantom protection despite the index being encapsulated by the storage service.

We introduce \proposed, a holistic index-and-buffer design optimized for storage-disaggregated OLTP with index pushdown. \proposed employs a tree-structured record buffer that unifies caching and buffering and accelerates both point and range queries. It further reduces the negative search penalty by caching negative search results. \proposed provides phantom protection that requires no knowledge of the storage-side data layout and is robust against large key domains, varying data skew and partition schemes. Experiments show that \proposed outperforms page-based \btree and the state-of-the-art \twotree by up to 11.42$\times$ and 2.97$\times$ on positive point queries, and achieves up to 4.76$\times$ speedup on negative searches over a hash-indexed record cache. For phantom protection, \proposed outperforms \pcl by up to 1.55$\times$ even under a dense key domain. Moreover, \proposed maintains stable performance while \pcl degrades sharply as the key domain grows large and sparse.
\end{abstract}



\received{17 April 2026}
\received[revised]{19 August 2026}
\received[accepted]{12 September 2026}

\maketitle

\section{Introduction} \label{sec:intro}

Modern cloud databases~\cite{verbitski2017amazon, socrates, zhou2021foundationdb, corbett12spanner, athena, snowflake} increasingly adopt a storage-disaggregated architecture, where compute and storage reside in separate services connected via a datacenter network. This architecture enables independent scaling and billing of compute and storage resources, improving elasticity and cost efficiency. However, compared to local disks, remote storage access incurs much higher latency and provides lower bandwidth due to network overhead.

Many cloud databases~\cite{verbitski2017amazon, socrates, PolarDB, Li2024GaussDB} directly adopt conventional indexing and buffer management techniques designed for local block devices: they maintain a monolithic B+tree and a buffer pool that manages index and data at page granularity. However, these designs are sub-optimal for disaggregated cloud storage: accessing remote storage incurs much higher latency than local devices and an index traversal introduces multiple serial round-trips over the network. Moving and caching data at page granularity also causes network traffic amplification and lowers cache efficiency. \looseness=-1

Another line of work, such as Deuteronomy~\cite{Levandoski2011DeuteronomyTS, levandoski15}, TiDB~\cite{tidb_serverless}, and Aurora DSQL~\cite{aurora_dsql}, leverages storage-side compute to push down the entire indexing function to the storage layer, and transfer only the hot data records instead of pages to the compute layer. Although this can eliminate network traffic amplification and avoid serial network round-trips, the existing index-pushdown approaches introduce three challenges under general OLTP workloads:

\noindent \circleb{1} \textbf{Inefficient Range Access}: Under index pushdown, compute-side caching is essential for avoiding remote round-trips on hot data. Existing solutions~\cite{lomet2018cost, rocksdb2021, levandoski15, gilad2020evendb, Treeline, anderson1998oracle, kanellisfaster} typically pair a hash-indexed record cache with a separate write buffer. While this significantly improves point-lookup performance, range queries still miss the cache because isolated cached records cannot reveal whether an interval is fully cached. The separate write buffer further causes read amplification and complicates consistency.

\noindent \circleb{2} \textbf{Negative Search Penalty}: Index pushdown with a record cache incurs significant performance degradation on negative searches, i.e., queries to keys that do not exist in the database. Each negative search must inevitably check both local cache and remote storage to confirm non-existence, causing unnecessary network round-trips. 

\noindent \circleb{3} \textbf{Phantom Protection}:
Phantom protection~\cite{berenson1995critique} is crucial for serializable isolation. Canonical next-key locking~\cite{mohan1989aries} needs to lock on the full index (i.e., B+tree) and therefore does not apply to storage disaggregation with index pushdown. Other physical-layout-agnostic methods are either impractical~\cite{predicate_lock,dibs} or less efficient~\cite{lomet2009locking}.

  \begin{table}[t]
    \caption{\proposed is the first to combine network-efficient index pushdown with full OLTP support.}
    \label{tab:comparison}
    \vspace{-0.08in}
    \centering
    \footnotesize
    \setlength{\tabcolsep}{3pt}
    \begin{tabular}{@{}lcccccc@{}}
      \toprule
      \textbf{Method} &
      \textbf{\shortstack[c]{No Serial\\Round-Trips}} &
      \textbf{\shortstack[c]{No Network\\Amplification}} &
      \textbf{\shortstack[c]{Cache\\Efficiency}} &
      \textbf{\shortstack[c]{Range\\Scan}} &
      \textbf{\shortstack[c]{Negative\\Search}} &
      \textbf{\shortstack[c]{Phantom\\Protection}} \\
      \midrule
      Page-Based \btree
        & \ding{55} & \ding{55} & \ding{55} & \cmark & \cmark & \cmark \\
      \midrule
      Index Pushdown w/ Record Cache
        & \cmark & \cmark & \cmark & \ding{55} & \ding{55} & \ding{55} \\
      \midrule
      \textbf{\proposed}
        & \textbf{\cmark} & \cmark & \cmark & \cmark & \cmark & \cmark \\
      \bottomrule
    \end{tabular}
  \end{table}

In this paper, we aim to answer a key question: \emph{How to design index and buffer management that fully optimizes for storage disaggregation?} Our answer is \proposed, a holistic index and buffer design that closes the gap in existing designs. \proposed combines the network efficiency of index pushdown with the full OLTP capabilities that page-based B+trees provide, as shown in Table~\ref{tab:comparison}. \proposed pushes down index traversal over the full dataset to the storage layer, while maintaining a tree-structured record buffer in the compute layer that caches hot data at record granularity. Data moves between the two layers through key-value interfaces. 

To address challenge \circleb{1}, we propose a \textbf{tree-structured record buffer}, an in-memory B+tree over hot cached records. It buffers updates and accelerates point and range queries. For range queries, we search the tree to check whether the entire range is cached; only if gaps are found do we issue a remote request for the minimal combined missing range. To reduce per-record metadata overhead, we approximate LRU by randomly traversing the tree, requiring only a 1-bit access flag per record. Recovery is simplified by logging only logical record updates, not tree structure changes. \looseness=-1

For \circleb{2} \textbf{negative search}, \proposed proactively caches \textbf{tombstone} entries in the tree-structured buffer to represent non-existent keys. For range queries, caching a continuous negative span as discrete tombstones is infeasible and space-inefficient; instead, \proposed only caches the negative range boundaries as tombstones. 

For \circleb{3} \textbf{phantom protection}, because index pushdown hides the global B+tree needed for classic next-key locking, \proposed uses the tree-structured record buffer as a logical lock table to protect the full key space. It locks the minimal \emph{cached} span covering the predicate; when the buffer is sparse or cold, this span can over-approximate the predicate (lock-scope amplification). To avoid this, \proposed first inserts lightweight \textbf{sentinel} records at uncached predicate boundaries, then locks the cached span. \looseness=-1

In summary, this paper makes the following contributions:
\begin{itemize}
    \item We design a tree-structured record buffer for index pushdown that unifies caching and write buffering at record granularity and accelerates both point and range queries.

    \item We develop a compact negative-caching mechanism that reduces remote probes for non-existent point and range keys.

    \item We introduce a phantom-protection mechanism that does not require exposing the storage layer's physical data layout or internal index, and remains robust across large key domains, varying skew, and partition schemes.

    \item We evaluate \proposed under storage disaggregation. It outperforms \btree and \twotree by up to 11.42$\times$ and 2.97$\times$ on positive point lookups, and beats hash-indexed record caches by up to 4.76$\times$ on negative searches. For phantom protection, it outperforms \pcl~\cite{lomet2009locking} by up to 1.55$\times$ and stays stable as the key domain grows.
\end{itemize}

\section{Background and Motivation}
\label{sec:background}

This section reviews storage disaggregation and existing index designs in cloud OLTP databases.

\begin{figure*}[t]
	\vspace{-0.1in}
        \centering
        \begin{subfigure}[b]{0.24\linewidth}%
            \centering
            \includegraphics[width=.95\linewidth]{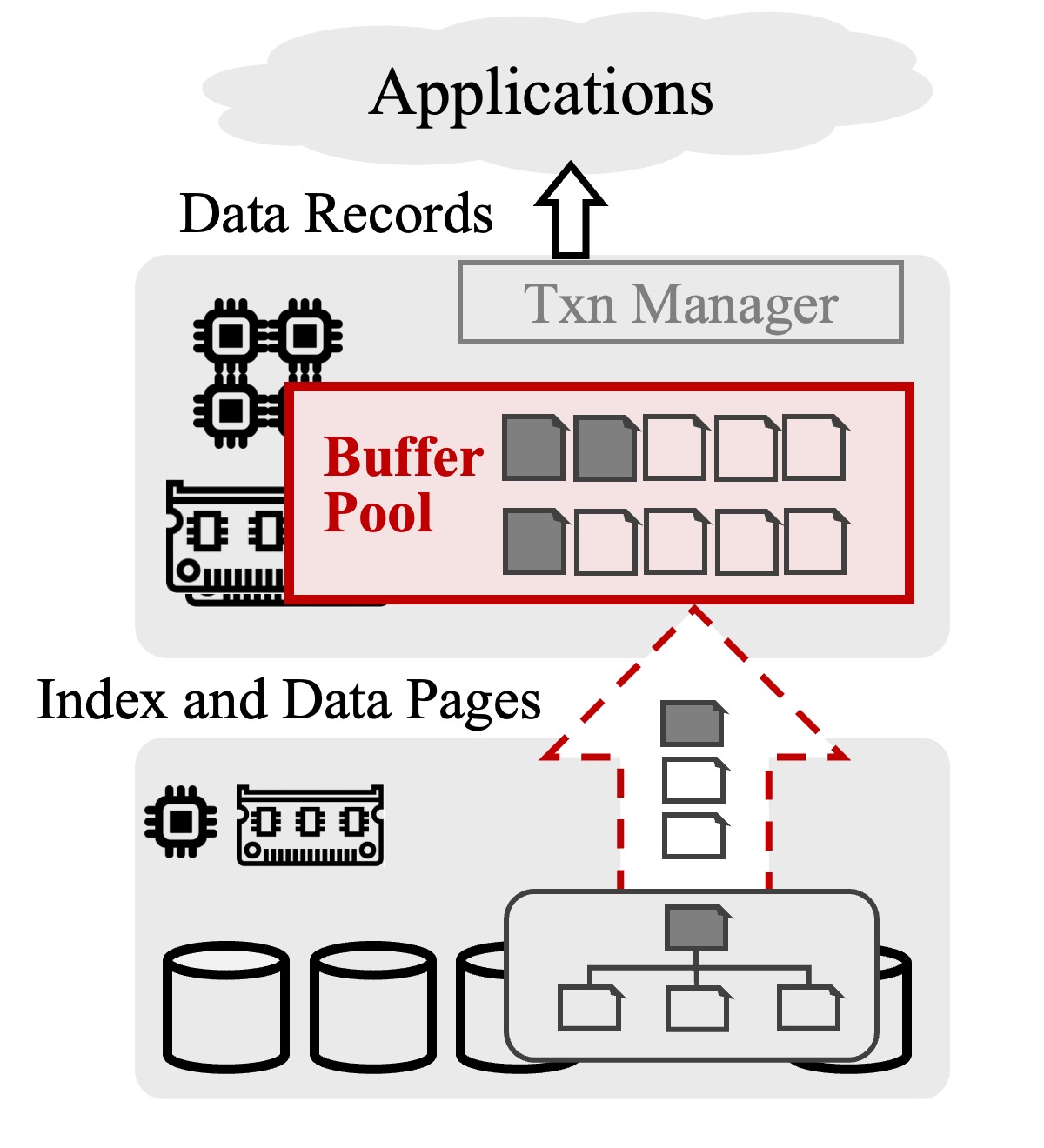}

            \caption{Page-Based (B+Tree)}

            \label{fig:bg-arch-page}
        \end{subfigure}\hfill
                \begin{subfigure}[b]{0.24\linewidth}%
            \centering
            \includegraphics[width=.95\linewidth]{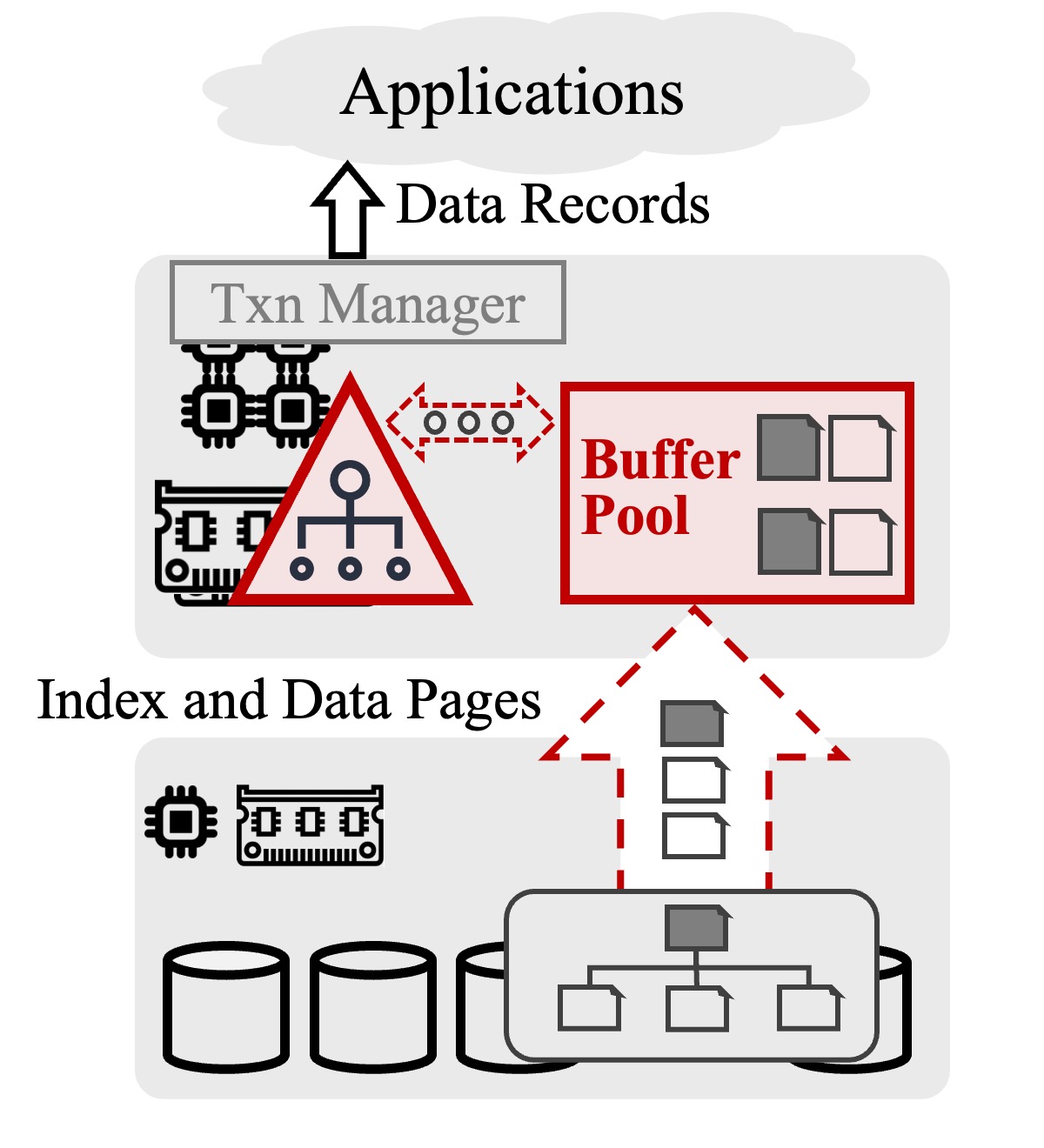}

            \caption{Page-Based (2-Tree)}

            \label{fig:bg-arch-2tree}
        \end{subfigure}\hfill
    \begin{subfigure}[b]{0.24\linewidth}%
    	\centering
        \includegraphics[width=.95\linewidth]{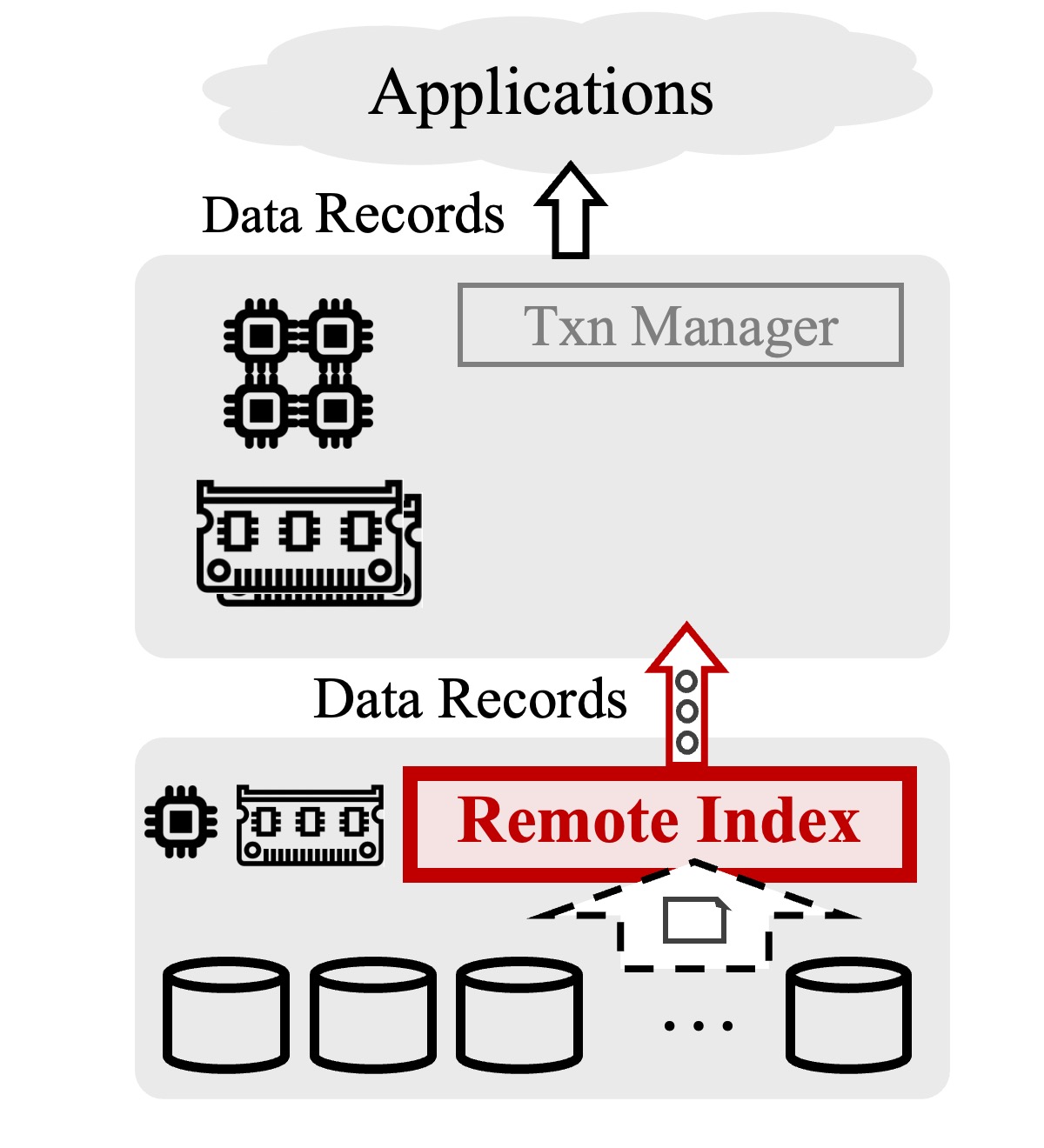}

        \caption{Remote-Only Index}

        \label{fig:bg-arch-remote}
    \end{subfigure}\hfill
    \begin{subfigure}[b]{0.24\linewidth}%
    	\centering
        \includegraphics[width=.95\linewidth]{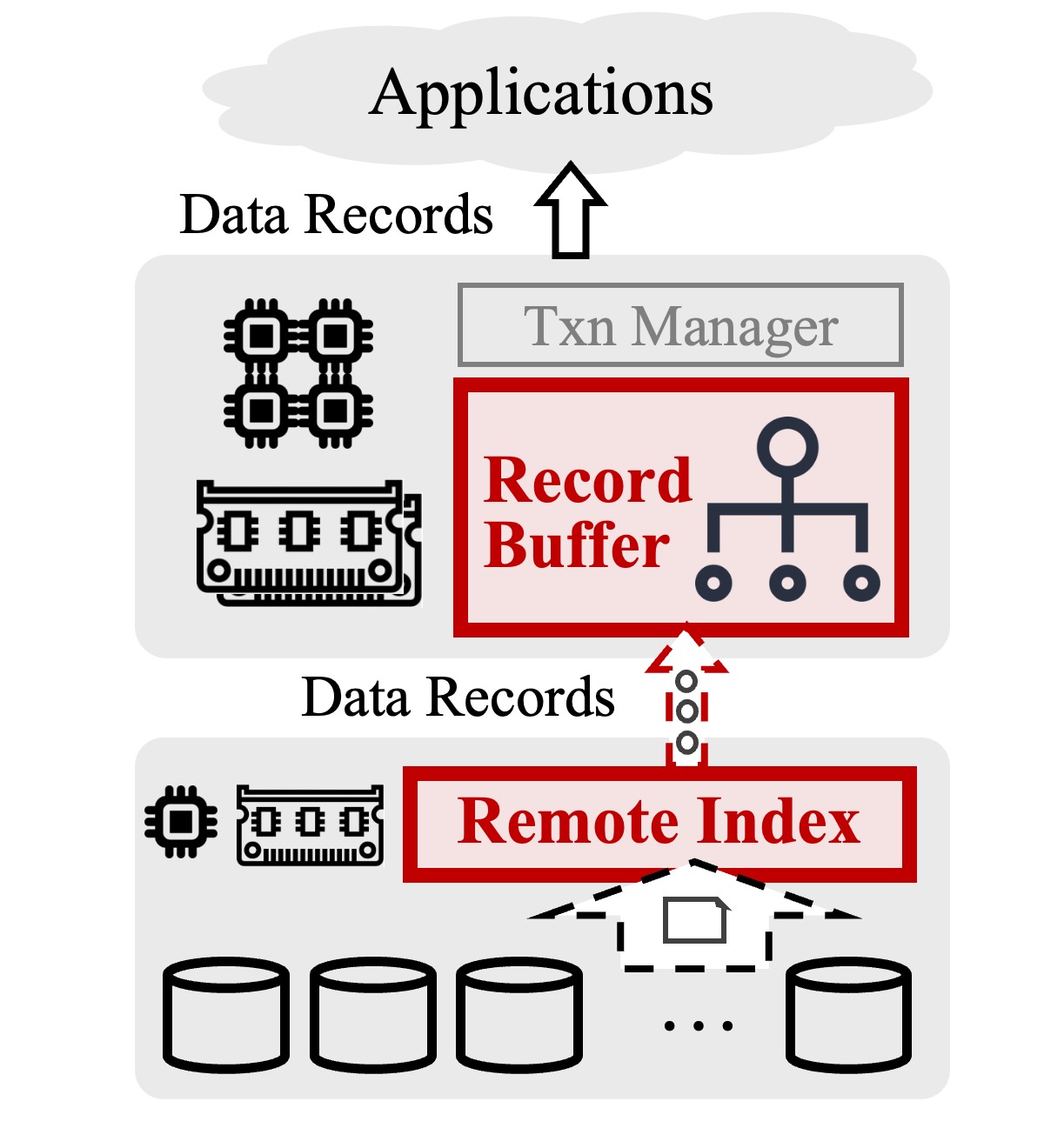}

        \caption{\proposed}

        \label{fig:bg-arch-nemo}
    \end{subfigure}%
      \vspace{-0.08in}
      \caption{\textbf{Index Architecture in Disaggregated DBs.} \normalfont{Arrow width shows transfer volume. Circles denote records; squares denote pages.}}
    \label{fig:bg_arch}
  \Description{Four side-by-side architecture diagrams, each with applications on top, a compute node with a transaction manager in the middle, and storage servers with disks at the bottom. (a) The compute node keeps a page buffer pool and pulls index and data pages over the network, drawn as a wide dashed arrow. (b) 2-Tree adds a hot record tree next to the page buffer pool but still pulls index and data pages. (c) The compute node keeps no cache; storage runs the remote index and returns only data records, drawn as a thin arrow. (d) Nemo keeps a tree-structured record buffer in the compute node on top of the remote index, so only records cross the network.}
\end{figure*}

\subsection{Storage Disaggregation}
\label{ssec:disaggregated storage}
Storage disaggregation replaces direct-attached disks with network-connected cloud storage services and has three key characteristics:

\boldtext{Network Latency and Bandwidth.} Remote storage access has far higher latency than local I/O: intra-region EC2-to-S3 round-trips take milliseconds, versus microseconds for NVMe SSDs. This gap amplifies cache-miss penalties, especially for serial I/Os (e.g., index traversal). Although top-tier instances offer raw bandwidth comparable to NVMe SSDs, mid-range instances are limited to 10--20~Gbps (1--2.5~GB/s), below a single NVMe SSD (3--7~GB/s), and protocol and software-stack overheads reduce it further. 

\boldtext{Fine-Grained Access.} Unlike local block devices restricted to fixed-size block I/O (e.g., 8~KB), network-connected storage communicates via high-level protocols (e.g., RPC or HTTP) and can expose APIs at finer granularity (e.g., records or bytes). For example, Azure Table Storage~\cite{azuretablestorage} and Google Cloud Datastore~\cite{google_cloud_datastore} provide key-value interfaces, and object stores such as Amazon S3~\cite{amazon_s3} and Azure Blob Storage~\cite{azure_blob_storage} support fine-grained \emph{range reads}.

\boldtext{Storage-Side Indexing.} Cloud storage services with local compute and memory can offer built-in indexing: Azure Table Storage~\cite{azuretablestorage} supports primary indexes; Google Cloud Datastore~\cite{google_cloud_datastore} automatically indexes each property; FoundationDB~\cite{zhou2021foundationdb} uses SQLite~\cite{sqlite} as disaggregated storage with internal B-tree indexes; Deuteronomy~\cite{Levandoski2011DeuteronomyTS}'s remote storage provides the Bw-tree~\cite{Levandoski2013bwtree}. 

\subsection{Page-Based Index}

\label{ssec:disk-index}

Conventional databases use page-based B+trees and buffer pools designed for local block devices~\cite{ubiquitousBtree, olson1999berkeley}; many cloud databases (Aurora~\cite{verbitski2017amazon}, Socrates~\cite{socrates}, PolarDB~\cite{PolarDB}, GaussDB~\cite{Li2024GaussDB}) apply them directly to disaggregated storage. These systems load index and data pages into the compute layer before processing queries (\figref{fig:bg-arch-page}): every cache miss triggers a high-latency network trip, and one index traversal causes multiple serial round-trips. Page-granularity movement also reduces cache efficiency and amplifies traffic: accessing one hot record transfers and caches its entire page.

Recent works~\cite{zhou2023two-tree,Siberia,anti-caching,Treeline} cache at record granularity. State-of-the-art \twotree~\cite{zhou2023two-tree} maintains a \textit{top tree} for hot records and a page-based \textit{bottom tree} for cold data, migrating hot tuples upward via a tuple-based protocol. These systems target local block devices: applied to disaggregation (\figref{fig:bg-arch-2tree}), \twotree still loads data and index pages over the network, retaining serial round-trips and traffic amplification. Dual trees also complicate parameter tuning (e.g., cache assignment) and transactional consistency, and \twotree lacks a full solution for phantom protection and negative search. \looseness=-1

\subsection{Remote-Only Index}
\label{ssec:remote-index}

Another line of work~\cite{Levandoski2011DeuteronomyTS, Treeline, zhou2021foundationdb} unbundles the database kernel into a logical transaction component and a data component that manages physical layout, communicating via record-oriented interfaces. The storage service manages indexes and returns only matching records (\figref{fig:bg-arch-remote}), avoiding page-transfer amplification and serial round-trips. This introduces three challenges, whose candidate solutions we review below:

\boldtext{Inefficient Range Access.} Some designs (e.g., Aurora DSQL~\cite{aurora_dsql}, TiDB~\cite{tidb_serverless}) forgo compute-side caching entirely, so every query pays a remote access. A natural remedy is a record cache for hot data~\cite{anderson1998oracle,Treeline}, which greatly improves point lookups; Deuteronomy~\cite{Levandoski2011DeuteronomyTS} pioneered this in disaggregated settings with a hash-indexed record cache. But record caches cannot serve range queries, since isolated records cannot reveal whether an interval is fully covered; and the cache is read-only, requiring a separate write buffer, which complicates concurrency control and forces reads to stitch across write buffer, cache, and remote storage. An effective design therefore requires a range-aware, update-in-place record buffer.

\boldtext{Negative Search.}
Under index pushdown, confirming non-existence requires remote probes. The cost is not limited to point lookups: most range queries are partially negative, needing to verify that no keys exist beyond those returned locally. Page-based indexes largely avoid this cost by caching index pages. Approximate membership structures target static or append-only sets and suit dynamic OLTP poorly: Bloom filters~\cite{bloom1970space} do not support ranges, handle deletions poorly, and suffer rising false positives under skew; practical range filters~\cite{zhang2018surf,vaidya2022snarf} struggle with variable-length keys and frequent updates; and auxiliary filters outside the index complicate transactional consistency and recovery. \looseness=-1

\boldtext{Phantom Protection.}
Conventional page-based indexes use next-key locking~\cite{mohan1989aries} for phantom protection, but it does not apply to remote-only indexes: the storage service hides the physical layout, so the compute layer must acquire locks before knowing which records fall within the requested range. Predicate locking~\cite{predicate_lock} decouples phantom protection from data layout, but its conflict detection is NP-hard and thus impractical. DIBS~\cite{dibs} supports only restricted predicates. Deuteronomy~\cite{Levandoski2011DeuteronomyTS} uses partition covering locks (\pcl)~\cite{lomet2009locking}, which lock logical partitions without requiring data layout. However, it is sensitive to partitioning and data distribution and cannot adapt to shifting skew. \pcl is one of our baselines (\secref{ssec:phantom_protection}). FoundationDB~\cite{zhou2021foundationdb} sidesteps phantoms via serializable snapshot isolation, which requires MVCC that not all storage services support and is orthogonal to lock-based serializability.

\section{System Model}
\label{sec:design}

As shown in \figref{fig:bg-arch-nemo}, \proposed pushes down the index and buffer pool to remote storage and caches only hot records at the compute layer in a tree-structured record buffer. This buffer is the basis for accelerating both point and range queries, mitigating negative-search overhead and enabling phantom protection without exposing storage internals. \proposed is the first to combine index pushdown with solutions to all three challenges in \secref{ssec:remote-index}.

\begin{figure}[t]
    \vspace{-0.1in}
    \centering
    \begin{minipage}[b]{0.32\linewidth}
        \centering
        \includegraphics[width=0.94\linewidth]{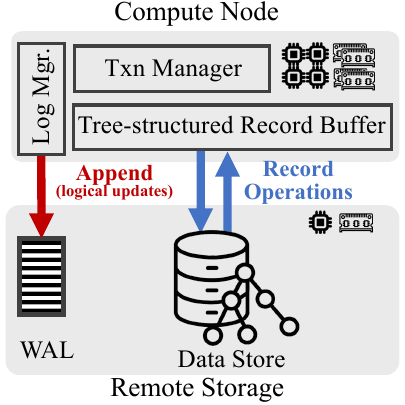}
        \vspace{-0.08in}
        \caption{\proposed System Model}
        \label{fig:flow}
        \Description{A compute node drawn above remote storage. The compute node contains a log manager, a transaction manager, and the tree-structured record buffer. A red arrow labeled Append (logical updates) runs from the log manager to the write-ahead log in remote storage, and blue arrows labeled Record Operations connect the record buffer with the data store.}
    \end{minipage}\hfill
    \begin{minipage}[b]{0.64\linewidth}
        \centering
        \includegraphics[width=\linewidth]{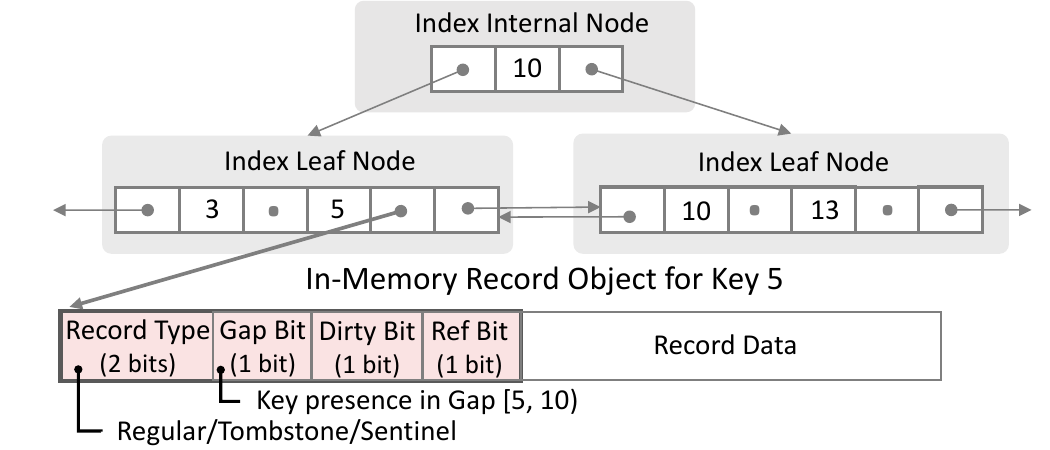}
        \vspace{-0.08in}
        \caption{Tree-Structured Record Buffer in \proposed}
        \label{fig:index}
        \Description{A small B+tree: an internal node with key 10 points to two linked leaf nodes holding keys 3 and 5, and keys 10 and 13. The leaf entry for key 5 points to its in-memory record object, which holds a 2-bit record type (regular, tombstone, or sentinel), a 1-bit gap bit telling whether keys may exist in the gap between 5 and 10, a dirty bit, a reference bit, and the record data.}
    \end{minipage}
\end{figure}

As shown in \figref{fig:flow}, database functionality is decomposed into two layers: a compute layer for transaction management and query execution, and a storage layer for data durability. For clarity, we present \proposed's design under a single database instance where the compute node has exclusive access to its storage. We discuss how \proposed maps to production cloud databases in \secref{sec:deployment}.

\boldtext{Compute Layer.} 
The compute layer (\figref{fig:flow}, top) has three components. The \textit{transaction manager} manages the transaction lifecycle and concurrency control. The \textit{log manager} logs transaction operations and ensures their durability. The \textit{tree-structured record buffer} caches hot data at record granularity and buffers updates. Only the updates from committed transactions are applied to the buffer. \looseness=-1

\boldtext{Storage Layer.} The disaggregated storage layer provides highly available, fault-tolerant persistence. It comprises two components (\figref{fig:flow}, bottom). \textit{Log store} persists the redo log of transaction operations in an append-only manner. \textit{Data store} maintains the database image. The data store is an atomic record store~\cite{rocksdb2021, mongodb}, which provides atomic operations on individual records. \tabref{tab:data_store_api} lists the minimal interfaces \proposed requires, which many cloud storage services satisfy (e.g., Azure Table Storage, Google Cloud Datastore).

\begin{table}[htbp]
        \caption{Data Store Atomic APIs}
        \label{tab:data_store_api}
        \vspace{-0.08in}
        \centering
        \footnotesize
        \begin{tabular}{@{}ll@{}}
        \toprule
        \textbf{Data Store API} & \textbf{Description} \\
        \midrule
        \textbf{Get(keys)} & Get one or a list of records \\
        \textbf{GetRange(sKey, eKey)} & Get records within a range [sKey, eKey] \\
        \textbf{Upsert(keys, values)} & Update one or a list of records; if absent, insert them instead \\
        \textbf{Delete(keys)} & Remove one or a list of records \\
        \bottomrule
        \end{tabular}
\end{table}

\proposed is independent of the data store's physical layout and internal index implementation, but it requires the four \textbf{atomic} record APIs in \tabref{tab:data_store_api}, including range access (\textit{GetRange}).  The data store handles its own recovery for durable atomic updates.

\section{\proposed Architecture}

\label{sec:indexprotocol}

This section presents the tree-structured record buffer (\secref{ssec:inmemoryindex}), how the core components integrate to support basic database operations (\secref{ssec:opertaions}), record-based buffer replacement (\secref{ssec:buf_replace}), and concurrency control (\secref{ssec:cc}).

\subsection{Tree-Structured Record Buffer}
\label{ssec:inmemoryindex}

The compute-layer \textbf{\textit{record buffer}} serves two roles: (i) buffering inserts, updates, and deletes, and (ii) caching hot records or ranges. It is the entry point for reads and writes: each query searches the buffer first, then decides whether to access remote storage. Operating over an incomplete local view requires reasoning about key-space intervals rather than individual keys: whether a range  is fully cached and where its holes lie, whether an interval is known to be empty, and which ranges must be locked for phantom protection. These tasks depend on the order and continuity of buffered keys, which an unordered cache (e.g., a hash table) does not expose. We therefore organize the buffer as a record-based B+tree (\figref{fig:index}) to accelerate point and range queries.
Key differences from page-based B+trees are:

\boldtext{Structural Design.} The record buffer employs an ephemeral index.
\begin{itemize}    
    \item \textit{Dynamic Construction}: Unlike page-based B+trees that maintain a persistent index over the full dataset, our tree is an ephemeral structure built only over cached data. It is built dynamically during cache activities and user modifications. Its footprint scales with cached data, avoiding a global index's memory overhead. As the index is not persisted, it is never fetched from storage, avoiding serial remote requests for traversal. Upon recovery, it is rebuilt on demand as the cache repopulates.
    
    \item \textit{Record Granularity}: The tree indexes records, not pages, eliminating buffer-pool page management and address translation. Each child pointer references an in-memory node or record. This pairs with our record-based buffer replacement so memory goes to hot records and network transfer stays at record granularity. 

\end{itemize}

A page-based \btree cannot match \proposed's benefits by simply shrinking its page size:  smaller pages reduce data cache and transfer amplification, but enlarge the index and raise its height, adding sequential remote round-trips to traversal. \proposed avoids this trade-off by pushing the global index to storage and building an ephemeral tree only over cached records, so its local index scales with cache size, not dataset size.

\boldtext{Per-Record Metadata.} Each buffered record carries 5 bits of metadata (\figref{fig:index}): a 2-bit record type (i.e., \textit{regular}, \textit{tombstone} and \textit{sentinel}), a 1-bit gap bit, and 2 buffer management bits (dirty and reference). Their roles are detailed later.

While we use a B+tree, the approach extends to other sorted trees like ART~\cite{leis2013art} or Masstree~\cite{mao2012masstree}. Since disaggregated databases are primarily I/O-bound, our contribution focuses on record-based buffering and record-granular transfer rather than the specific tree. We leave a systematic comparison to future work.

\subsection{Database Basic Operations}
\label{ssec:opertaions}

This section outlines how standard database operations use the core components (record buffer and remote storage). To focus on component interactions, we omit buffer replacement internals, latches, and tree-internal procedures in \algoref{alg:operations}. 

\begin{algorithm}[t]

    \SetKwProg{myfun}{Function}{}{}
    \DontPrintSemicolon

    \myfun{DB.LookUp(key)}{ 
        \(\textit{st}, \textit{rec} \gets \textit{Buffer.Search}(key)\) \;
        \lIf{\(\textit{st} = \texttt{Succeeded}\)}{\Return \textit{rec}}
        \lElseIf{\(\textit{st} = \texttt{KeyNotExist}\)}{\Return null}
        \ElseIf {\(\textit{st} = \texttt{BufferNotFound}\)} {
            \(\textit{admit\_st}, \textit{remote\_rec} \gets \textit{\shl{Admit(key)}}\) \;
            \Return \((\textit{admit\_st}=\texttt{Succeeded})?\,\textit{remote\_rec}:\,\textit{null}\)
        }
    }

    \myfun{DB.Insert(record)}{
       \(\textit{rec} \gets \textit{DB.LookUp}(\textit{record.key})\) \;
       \lIf{\(\textit{rec} \neq \texttt{null}\)}{\Return \texttt{KeyAlreadyExist}}
       \Return \(\textit{\shl{Buffer.Insert(record,\ mode\ =\ \texttt{UserOps})}}\)
    }

    \myfun{DB.Delete(key)}{
       \(\textit{rec} \gets \textit{DB.LookUp}(\texttt{key})\) \;
       \lIf{\(\textit{rec} = \texttt{null}\)}{\Return \texttt{KeyNotExist}}
       \(\textit{\shl{Buffer.Insert(MakeTombstone(key),\ mode\ =\ \texttt{UserOps})}}\) \;
       \Return \texttt{Succeeded}
    }

   \myfun{DB.RangeScan(start, end)}{%
      \(\textit{recs}, \textit{hole} \gets \textit{Buffer.RangeScan}(start, end)\) \;
      \lIf{\(hole=\varnothing\)}{
        \Return\(\textit{recs}\)
      }
      \(\textit{st}, \textit{remote\_recs} \gets \textit{\shl{AdmitRange(hole)}}\) \;
      \Return \((\textit{st}=\texttt{Succeeded})?\,\textit{remote\_recs} \cup \textit{recs}:\,\textit{recs}\)
   }
    \caption{\textbf{Database User Operations}. Record buffer modifications are marked in \shl{gray}.}
    \label{alg:operations}
\end{algorithm}

Read operations first probe the record buffer, then trigger cache admission for the missing records. Write operations first check key existence, then insert the record into the record buffer. Dirty records are flushed to remote storage during cache eviction. As highlighted in \algoref{alg:operations}, the record buffer can be modified by both user operations (insert/delete) and cache activities (admission/eviction).

\subsubsection{LookUp} A lookup first probes the record buffer (line 2). If found, it returns immediately (line 3). If a gap bit or tombstone shows the key does not exist, it returns null (line 4). Otherwise, we fetch the record from remote storage and admit it if needed (lines 5--6). 

\subsubsection{Insert} An insert first checks via a Lookup whether the key exists (line 9). If found, we return a KeyAlreadyExist error (line 10). Otherwise, we insert the record into the record buffer (line 11). 

\subsubsection{Delete} A delete is similar to an insert. The difference is that it inserts a tombstone rather than a regular record (line 15). The tombstone is written back during eviction via the remote storage's delete operation (\secref{ssec:buf_replace}).

\subsubsection{Range Scan} A range scan searches the record buffer to find all cached records within the requested range, and any missing hole via the gap bits (line 18). Multiple holes are coalesced into one minimal covering hole. If none exists, cached records are returned directly (line 19); otherwise, the missing range is fetched from remote storage (line 20) and merged with cached records.

\subsection{Buffer Replacement}
\label{ssec:buf_replace}

Managing buffers at record granularity presents two challenges: (i) tracking per-record recency and selecting victims with minimal per-record space overhead, and (ii) evicting records safely in batches with parallelism. Existing solutions incur high overhead. For example, a record-level CLOCK~\cite{Corbato1969Paging} requires a per-record clock-hand pointer. Moreover, it forms a far larger circle than page-based CLOCK, slowing each sweep and raising the need for batched and parallel eviction. Yet its sequential victim scan limits parallelism.

\proposed approximates LRU using probabilistic tree traversals, avoiding heavy per-record metadata (e.g., LRU lists or clock hands). It adds only two bits per record: a \textit{dirty} bit for modifications and a \textit{reference} bit for access since the last probe. Each eviction performs a random tree traversal to select a leaf; with uniform leaf sampling, the expected interval to revisit a leaf statistically resembles the circular sweep of CLOCK. This design naturally supports parallel, batched eviction through concurrent tree traversals. In \algoref{alg:buffer}, all buffer operations are implemented as tree manipulations. \looseness=-1

\begin{algorithm}[t]
    \SetKwProg{myfun}{Function}{}{}
    \DontPrintSemicolon
  
    \myfun{Admit(key)}{%
    \(r \gets \textit{remote\_storage.Get}(key)\) \;
    \(\textit{rec} \gets (r=\textit{null})?\,\textit{MakeTombstone}(key):\,r\) \;

    wait \(\textit{AvailableBufferSpace}()\) \;
    \textit{\shl{Buffer.Insert(rec, mode = \texttt{CacheSingle})}} \;
    \Return \((\textit{r}=\textit{null})?(\texttt{KeyNotExist},\textit{r}):(\texttt{Succeeded},\textit{r})\)
  }

    \myfun{AdmitRange(start, end)}{%
      \(\textit{res} \gets \textit{remote\_storage.GetRange}(start, end)\) \;
      \lIf{\(start \notin \textit{res}\)}{
        \textit{res.addFront(MakeTombstone(start))}
      }
      \lIf{\(end \notin \textit{res}\)}{
        \textit{res.addBack(MakeTombstone(end))}
      }
      wait \(\textit{AvailableBufferSpace}()\) \;
      \lForEach{\(r \  \textit{in} \ \textit{res}\)}{
        \(\textit{\shl{Buffer.Insert(r, mode = \texttt{CacheRange})}}\)
      }
      \Return \((\textit{res}=\varnothing)?(\texttt{NoKeyExist}, \textit{null}):(\texttt{Succeeded}, \textit{res})\)
    }

    \myfun{Evict()}{%
    \(\textit{leaf} \gets \textit{Buffer.RandomTopDownTraverse}()\) \;
    \(\textit{evicts}, \textit{upserts}, \textit{deletes} \gets \varnothing\) \;
    \ForEach{\(rec \  \textit{in} \  \textit{leaf}\)}{
      \uIf{\(rec.\textit{ref\_bit}=0\)}{
        \textit{evicts.add(rec)} \;
        \uIf{\(\textit{rec.isDirty}()\)}{
            \lIf{\textit{rec.isTombstone()}}{
                \textit{deletes.add(rec)}
            }
            \lElseIf{\textit{rec.isRegular()}}{
                \textit{upserts.add(rec)}
            }            
            }
      }
      \lElse{
        \(rec.\textit{ref\_bit} \gets 0\)  %
      }
    }
    \(\textit{remote\_storage.Upsert}(\textit{upserts})\) \;
    \(\textit{remote\_storage.Delete}(\textit{deletes})\) \;
    \(\textit{\shl{Buffer.Delete(evicts)}}\) \;
    \Return \texttt{Succeeded}
  }

    \caption{\textbf{Buffer Replacement}. Record buffer modifications are marked in \shl{gray}.}
    \label{alg:buffer}
\end{algorithm}

\subsubsection{Cache Admission} When a user-triggered search misses the record buffer, \emph{cache admission} is invoked (lines 6, 20 in \algoref{alg:operations}) to fetch the records from remote storage (lines 2, 8 in \algoref{alg:buffer}). When sufficient buffer space is available (lines 4, 11), the fetched records are inserted into the buffer (lines 5, 12) and the result is returned. Range admission differs from single-key admission: the system “fences” the range by caching bounds if absent as tombstone entries (lines 9–10). These boundary tombstones, together with the gap bits, enable efficient negative range searches (\secref{sec:negativesearch}). \looseness=-1

For simplicity, Algorithm~\ref{alg:buffer} admits every miss into the buffer. In our evaluation, we fully warm the cache under steady workloads before each experiment, so it does not suffer cold-start thrashing after warm-up. In practice, admission can be more selective (e.g., probabilistic) to avoid flooding the cache with replacement traffic; we leave a systematic study of selective admission to future work.

\subsubsection{Cache Eviction} Eviction reclaims space by removing least recently used records from the buffer. Background tasks periodically perform eviction: each randomly selects a leaf node via a top-down traversal of the tree-structured buffer (line 15), then scans the leaf to collect victims and reset reference bits (lines 17–23). Records with a reference bit of 0 are evicted (line 19); following a second-chance policy, records with a reference bit of 1 are retained and reset (line 23). Reference bits thus cool exactly as in CLOCK: set on access and cleared by each eviction pass, so a record is reclaimed by the next pass that reaches it unless re-accessed in between. Dirty victims, whether tombstones or regular records, are batched and written back to storage (lines 24–25). The selected victims are then deleted from the buffer in batch (line 26). We run eviction tasks from multiple threads for parallelism. \looseness=-1

Eviction never scans the whole tree: each task follows one root-to-leaf path and reclaims a victim batch from the selected leaf, costing $O(\log_b n)$ for $n$ cached records with fanout $b$, not $O(n)$. With large fanout, tens of millions of records form a tree of 3--4 levels, so each batch touches few nodes. \looseness=-1

\subsection{Concurrency Control}
\label{ssec:cc}

Similar to traditional disk-based databases, we deal with transactional concurrency control and concurrent tree structure operations separately. For transactional concurrency control, we use 2PL to ensure isolation. All transactions follow serializable isolation through the \textit{NO\_WAIT} protocol, which avoids deadlocks. Concurrency control runs entirely in the compute layer: each partition is exclusively owned by a single compute node (\secref{ssec:partitioned_writer}), so all conflicting accesses to a record meet at its owner, which acquires locks on buffered records and holds them until commit or abort. The storage layer performs no transactional locking, providing only atomic record operations (\tabref{tab:data_store_api}) and a durable WAL (\secref{sec:recovery}); cross-partition transactions use standard 2PC~\cite{gray1978notes}. Modifications are applied to the record buffer only after commit. Phantom protection is discussed in \secref{sec:phantomprotection}. For concurrent tree operations on the buffer, we use latches and lock coupling for correctness.

Note that, unlike conventional page-based indexes where the index is modified only by user operations, modifications (insertions/deletions) on the tree-structured buffer in \proposed can be initiated by either user requests or cache activities (admissions/evictions). Only user-triggered modifications acquire transactional-lifespan locks, while cache-triggered modifications acquire only brief latches. \looseness=-1

\section{Negative Search}
\label{sec:negativesearch}

Negative searches query keys that do not exist in the database. As discussed in \secref{ssec:remote-index}, negative searches are expensive under index pushdown with a record-granular cache because the record buffer cannot distinguish non-existent keys from uncached ones, forcing remote probes. Existing approximate membership structures (e.g., Bloom filters) fit poorly with dynamic OLTP workloads, and maintaining such auxiliary structures complicates concurrency control and recovery. We seek a lightweight solution with the following properties: (1) \textbf{Volatile}: avoid external data structures that must be continuously persisted or pre-computed, or that intrude on the decoupled storage layer; (2) \textbf{Safety}: allow false positives but not false negatives; (3) \textbf{Adaptivity}: react quickly to shifting access skew; (4) \textbf{Tiny Space}: incur minimal memory overhead.

\begin{table}[t]
        \caption{\textbf{Record Metadata Update Rules} \textnormal{(If \textit{rec.key} is found in the buffer, cursor is the matching record; otherwise, cursor is its predecessor)}}
        \label{tab:meta_buf_insert}
        \vspace{-0.08in}
        \centering
        \small
        \setlength{\tabcolsep}{1.2pt}
        \setlength{\aboverulesep}{0pt}
        \setlength{\belowrulesep}{0pt}
        \begin{tabularx}{\linewidth}{|>{\raggedright\arraybackslash}p{51.5pt}|
                        >{\raggedright\arraybackslash\hsize=1.0874\hsize}X|
                        >{\centering\arraybackslash\hsize=1.2053\hsize}X|
                        >{\centering\arraybackslash\hsize=0.7074\hsize}X|
                        >{\centering\arraybackslash}p{22pt}|}
        \hline
        \multirow{2}{=}{\textbf{User/Cache Ops}} &
        \multirow{2}{*}{\textbf{Buffer Ops}} &
        \multicolumn{3}{c|}{\textbf{Record Metadata After Operation}} \\ \cline{3-5}
        & & \textbf{Record Type} & \textbf{GapBit} & \textbf{Dirty} \\ \midrule
        DB.Insert() & Buffer.Insert(rec, UserOps) & rec.type (i.e., Regular) & cursor.gap\_bit & true \\ \hline
        DB.Delete() & Buffer.Insert(rec, UserOps) & rec.type (i.e., Tombstone) & cursor.gap\_bit & true \\ \hline
        DB.Admit() & Buffer.Insert(rec, CacheSingle) & rec.type (i.e., Regular\slash Tombstone) & 1 & false \\ \hline
        DB.\allowbreak AdmitRange() & Buffer.Insert(rec, CacheRange) & rec.type (i.e., Regular\slash Tombstone) & 1, if rec is the range end; 0, otherwise & false \\ \hline
        DB.Evict() & Buffer.Delete(rec) &
        \multicolumn{3}{>{\centering\arraybackslash\hsize=\dimexpr 1.9126\hsize+22pt+4\tabcolsep+2\arrayrulewidth\relax}X|}{%
        Delete the record; if it is a regular record, update its predecessor’s gap bit to 1} \\ \hline
        \end{tabularx}
\end{table}

\subsection{Negative Caching}

An intuitive approach is to cache negative search results in the buffer. Concretely, on a negative lookup, we insert a tombstone record into the buffer to mark an absent key, so subsequent lookups for the same key hit the cached tombstone and avoid a remote request. While \proposed uses this method for point lookups, it does not apply to range searches: an empty (or partially empty) range may cover a vast or even unbounded key space. Caching all empty spans as discrete tombstones consumes substantial cache space, especially for large-domain key types such as strings.

\proposed adds a \textbf{Gap Bit} to the metadata of each buffered record (either regular or tombstone). It encodes the interval state between the current record and its immediate successor in the buffer: (1)~$GapBit = 0$: the interval is \textit{confirmed empty} in remote storage; (2)~$GapBit = 1$: the interval state is \textit{unknown}, and searches in this gap must probe remote storage to confirm the existence of keys.

\proposed combines tombstones with gap bits to cache negative results compactly. For a negative point lookup, it simply inserts a tombstone. \proposed represents a completely cached range using ordered buffer records and gap bits together, rather than gap bits alone. Boundary records, either regular or tombstone, delimit the range, while gap bits of 0 between consecutive records confirm complete coverage of the intervening intervals. For example, in a negative range query (e.g., $[15,20]$), it caches the boundary keys 15 and 20 as tombstones and sets the gap bit of record 15 to 0. This effectively caches the negative range, allowing future queries for any sub-range (e.g., $[16, 18]$) to be pruned locally. This mechanism also handles partially empty ranges: if 15 exists but no keys exist in $(15,20]$, the buffer caches 15 as a regular record with its gap bit $=0$ and caches 20 as a tombstone. 

\begin{algorithm}[htbp]
        \SetKwProg{myfun}{Function}{}{}
        \DontPrintSemicolon
           
        \myfun{Buffer.Search(key)\quad\codeComment{Called by DB.LookUp()}}{%
        \nonl \codeComment{If found, cursor is the matching record; otherwise, cursor is the key's predecessor record.}\;
        \(\textit{found}, \textit{cursor} \gets \textit{tree.Search(key)}\) \;
        \uIf{found}{
            \SetKwIF{If}{ElseIf}{Else}{\textbf{if}}{}{\textbf{else if}}{\textbf{else}}{}
            \Return \(\textit{cursor.isTombstone()}?(\texttt{KeyNotExist}, cursor): (\texttt{Succeeded}, cursor)\) 
        }
        \lElse{
        \Return \((\textit{cursor.gap\_bit}=0)?(\texttt{KeyNotExist},\textit{null}):(\texttt{BufferNotFound},\textit{null})\)
        }
      }
    
      \nonl\codeComment{Called by DB.Insert(), DB.Delete(), DB.Admit(), DB.AdmitRange()}\;
      \myfun{Buffer.Insert(rec, mode)}{
        \nonl\codeComment{INSERT acts as UPSERT because tombstones exist}\;
        \(\textit{found}, \textit{cursor} \gets \textit{tree.Search}(rec.key)\)\;
        
        \nonl \codeComment{generate the record metadata according to \tabref{tab:meta_buf_insert}} \; 
        \(meta \gets \textit{GenRecMeta(rec, mode, cursor)}\) \;       

        \lIf{\textit{found}}{
            \textit{cursor.Update(meta, rec.value)}
        }\lElse{
            \textit{tree.Insert(MakeRecord(rec.key, meta, rec.value))}
        }

        \Return \texttt{Succeeded}
      }
       
        \myfun{Buffer.Delete(rec)\quad\codeComment{Only called by DB.Evict()}} {
           \(\textit{prev\_record} \gets \textit{tree.Delete(rec.key)}\)\;
           \lIf{\textit{rec.isRegular()}} {
              \textit{prev\_record.Update(gap\_bit = 1)}
           }
           \Return \texttt{Succeeded}
        }
        
        \caption{\textbf{Buffer Operations with Negative Caching}.}
 
        \label{alg:wrappers}
    \end{algorithm}

\subsection{Buffer Operations with Negative Caching}

Following \algoref{alg:wrappers}, we describe how negative caching reduces remote probes (\textit{Buffer Search}) and how buffer modifications preserve the tombstone and gap-bit invariants (\textit{Buffer Insert/Delete}).

\boldtext{Buffer Search.} User reads (\textit{DB.LookUp} and \textit{DB.RangeScan} in \algoref{alg:operations}) begin by searching the buffer. If the target key or range is not fully resolved locally, the system probes remote storage. \textit{Buffer.Search} follows standard B+tree traversal but avoids remote probes in two cases: if it encounters a cached tombstone (line 4), or if the target is absent but its predecessor's gap bit is 0 (line 5). Both immediately return \textit{KeyNotExist}. Similarly, \textit{Buffer.RangeScan} aggregates cached records and identifies holes via gap bits of 1. If none are found, it returns the results without remote access.

\boldtext{Buffer Insert.} Buffer insertion acts as an UPSERT since the target key may already exist (e.g., as a cached tombstone from prior negative searches). Following \algoref{alg:wrappers}, it first searches for the key (line 7) and derives metadata using the state-transition rules in \tabref{tab:meta_buf_insert} (line 8). If the key is found, it updates the record with the new value and metadata (line 9); otherwise, it inserts a new record (line 10). \textit{Buffer.Insert} is triggered in four scenarios (\tabref{tab:meta_buf_insert}); we explain how the metadata is determined: \looseness=-1

\begin{itemize}
    \item \textit{User Insert}: The inserted record must be a regular record. It inherits the gap bit from the cached record (if present) or the key's predecessor. The dirty bit is true as the record is modified. \looseness=-1

    \item \textit{User Delete}: Instead of physically removing the record, a user delete inserts a tombstone record. The gap bit and dirty bit follow the same rules as a user insert.

    \item \textit{Cache Admit (Point)}: The record type is \textit{Regular} or \textit{Tombstone}, depending on the remote probe result. The gap bit is conservatively set to 1, as neighbor adjacency is unknown. The dirty bit is \textit{false}, as admission does not modify the record.

    \item \textit{Cache Admit (Range)}: Range-admitted records can be \textit{Regular} or \textit{Tombstone}, because the range query caches positive records as regular records and negative range boundaries as tombstones. Since admitted ranges are contiguous, \proposed sets the gap bit to 0 for all records except the final one, sealing internal gaps. The final record (range end) receives a gap bit of 1, as the state beyond the range is unknown. The dirty bit is \textit{false}.
\end{itemize}

\boldtext{Buffer Delete.} Buffer deletion is only triggered during cache eviction. It removes the record (line 13). If the deleted record is regular, its predecessor's gap bit is reset to 1 (line 14) as the record now resides only in remote storage and must be probed remotely.

\subsection{Eviction for Negative Caching}
As described in \secref{ssec:buf_replace}, \proposed uses a reference bit to track the access recency of each cached record and uses it to guide the choice of eviction candidates. Since \proposed caches negative search results within the record buffer, we must also track ``\textbf{negative hits}'' to prevent hot negative information from being evicted. We use the following rules to track the access recency of negative information (tombstones and gap bits). If a search accesses a tombstone, we set the tombstone's reference bit to 1. If a search falls into a confirmed empty interval (i.e., utilizing a gap bit of 0 to prune a remote probe), we treat the gap as accessed by setting the reference bits of both neighbor records framing the gap (the predecessor and successor) to 1. This ensures that the records defining a hot empty range are preserved in the buffer. The neighbor records may be either regular records or tombstones. Conversely, a tombstone or gap boundary that is not referenced again before the next sweep of its leaf is evicted like any other cold record, so negative information that is never reused cannot accumulate beyond one eviction pass.

\boldtext{Future Optimizations.} The costs of building and evicting gap-bit-0 coverage are asymmetric: building coverage requires a range admission that fetches the interval from storage (\algoref{alg:buffer}, lines 7--13), whereas evicting a single regular record resets its predecessor's gap bit to 1 (\algoref{alg:wrappers}, line 14) and invalidates the interval it covered. The \textbf{negative hits} rules above mitigate this asymmetry: each search that exploits a gap bit of 0 refreshes the reference bits of both framing records, so hot coverage is rarely evicted. However, under sustained memory pressure, eviction can outpace these refreshes and reduce the benefit of negative caching. As future work, a cost-aware replacement policy can grant gap-boundary records higher retention priority, reflecting their higher reconstruction cost. \looseness=-1

\section{Phantom Protection}
\label{sec:phantomprotection}

We first discuss the limitations of prior approaches, then present \proposed{}’s approach with an example.

\subsection{Phantom Protection under Index Pushdown}
\label{ssec:pp-prior}

Phantom protection prevents the ``phantom read'' anomaly, where a transaction re-executes a predicate and unexpectedly sees different rows due to concurrent inserts or deletes. This is particularly challenging under index pushdown, where storage encapsulates the physical data layout and the compute layer sees only a cached subset of records. Existing solutions either do not apply here or sacrifice efficiency. We review two representative approaches to motivate the need for an efficient, storage-agnostic solution.

\noindent \textbf{Next-key locking}~\cite{mohan1989aries} locks each indexed record in the requested range and the negative gap following it. However, it requires the full-key-space ordered index in the compute layer to navigate entries and gaps. It cannot be directly applied to index pushdown designs.

\noindent
\textbf{Partition Covering Lock (PCL)}~\cite{lomet2009locking} is a storage-agnostic solution that logically partitions the key space into ranges and locks the minimal set of partitions covering the requested predicate range. However, PCL locks coarsely and is highly sensitive to the partitioning strategy. Static partitions suffer under skew when hot keys cluster in a few partitions, while dynamic re-partitioning is costly and lagging. Moreover, for vast, sparse key domains (e.g., URLs), keeping the partition count manageable forces wide partition ranges that introduce excessive false conflicts.

\subsection{Phantom Protection in \proposed}
\label{ssec:nextkey}

\proposed achieves storage-agnostic phantom protection by adapting next-key locking to the cached hot records in the compute layer.

\boldtext{Locking Buffer Ranges.} Conventional next-key locking relies on the full index to lock precise keys and the exact negative gaps between them. In contrast, \proposed treats the tree-structured record buffer as a sorted logical lock table and locks the minimal \emph{cached} span that encloses the requested predicate. Concretely, it traverses the B+tree-indexed buffer to find the largest cached key $k_l \le \text{query}_{start}$, then acquires next-key locks in key order until reaching the first cached key $k_r > \text{query}_{end}$. This is storage-agnostic, partition-free, and adapts to shifting skew since the locked span is defined by the cached hot set, which evolves with query hotspots.

However, when the buffer is sparse (e.g., small or cold), the gaps between cached keys may be much larger than the true negative gaps in storage, forcing \proposed to lock a wider enclosing cached span than the predicate range. For example, for predicate $[5,10]$, if the buffer contains only keys 1 and 100 (while 5 and 10 reside in storage), \proposed must conservatively lock the cached span $[1,100]$.

\boldtext{Sentinel Optimization.}
To tighten lock scope, \proposed introduces \textit{\textbf{sentinel}} records: lightweight placeholders (key + metadata, no payload) for keys not yet observed in the buffer. If a predicate boundary key is uncached, \proposed inserts a sentinel at that boundary as an anchor to lock the range precisely. A range query executes as follows:

\begin{enumerate}
    \item \textbf{Sentinel Insertion:} Insert sentinels for $\text{query}_{start}$ and $\text{query}_{end}$ if not cached in the buffer (e.g., keys 5 and 10).
    \item \textbf{Range Locking:} Acquire next-key locks over the minimal cached span (e.g., [5, 10]) covering the requested range.
    \item \textbf{Resolution:} After fetching results from storage, resolve sentinels during cache admission: if the key exists, convert it into a regular record; otherwise, convert it into a \textit{tombstone}.

\end{enumerate}

\boldtext{Sentinel Overhead.}
Sentinels are lightweight by construction. Each range query inserts at most two sentinels, one per uncached predicate boundary, negligible next to the fetched records the query already admits into the buffer. Each sentinel holds only a key and 5 bits of metadata, with no payload, and lives briefly: at cache admission it resolves into a regular record or a tombstone that serves subsequent queries under the normal replacement policy; unresolved sentinels thus never outnumber the in-flight range queries. An extreme flood of tiny, random negative range queries could still accumulate cold tombstones, which are gradually reclaimed by normal eviction. As future work, background garbage collection could activate under tombstone pressure, coalescing nearby tombstones into gap-bit-0 coverage via large range scans.    

\subsection{Example}
\label{ssec:phantomexample}

\begin{figure}[t]
    \vspace{-0.1in}
    \centering
    \begin{subfigure}[t]{0.62\linewidth}%
    	\centering
        \includegraphics[width=\linewidth]{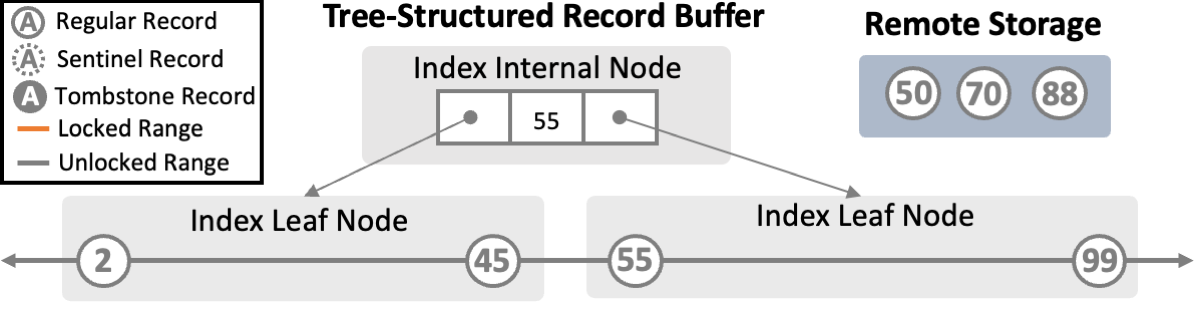}        
     \caption{Initial State}
    \label{fig:illus1}
    \end{subfigure}

    \begin{subfigure}[t]{0.62\linewidth}%
    	\centering
        \includegraphics[width=\linewidth]{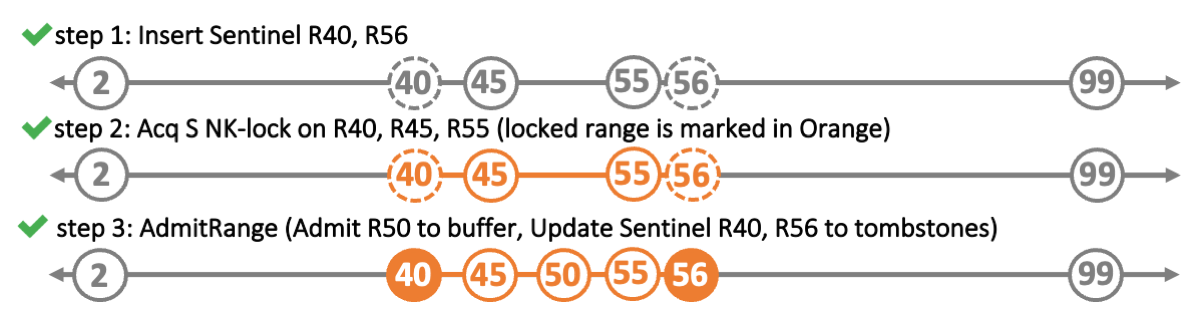}        
     \caption{T1: RangeScan(40, 56)}
    \label{fig:illus2}
    \end{subfigure}

    \begin{subfigure}[t]{0.62\linewidth}%
    	\centering
        \includegraphics[width=\linewidth]{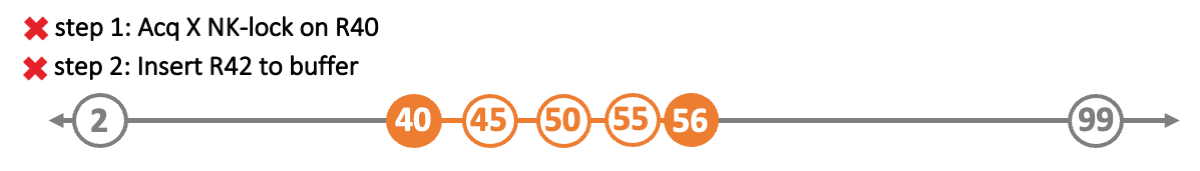}

        \caption{T2: Insert(42, ``Record Data'')}
    
        \label{fig:pp_q2}
    \end{subfigure}

    \begin{subfigure}[t]{0.62\linewidth}%
    	\centering

        \includegraphics[width=\linewidth]{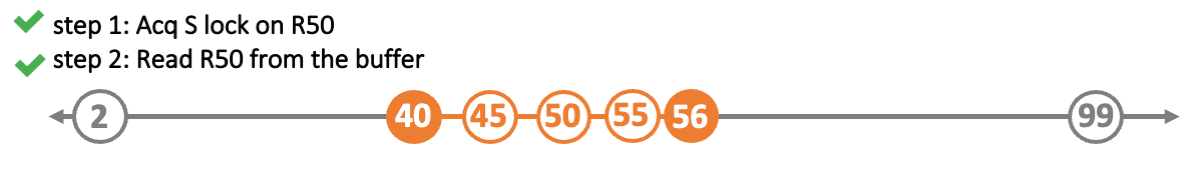}
    
        \caption{T3: LookUp(50)}
    
        \label{fig:pp_q3}
    \end{subfigure}%

    \vspace{-0.08in}
    \caption{Example. \textnormal{Locked scope is highlighted in orange. T2 and T3 execute when T1 gets to step 3.}}

    \label{fig:phantomillustration}
  \Description{Four stacked diagrams of one scenario. Each shows the record buffer as a tree over a key line from 2 to 99, next to remote storage holding keys 50, 70, and 88. (a) The buffer caches records 2, 45, 55, and 99. (b) T1 inserts sentinels at 40 and 56, takes shared next-key locks on 40, 45, and 55 (drawn in orange), then admits record 50 and turns the sentinels into tombstones. (c) T2 fails to take an exclusive next-key lock on 40, so it cannot insert 42. (d) T3 takes a shared lock on 50 and reads it from the buffer.}
\end{figure}

We use a concrete example to illustrate \proposed's phantom protection. Consider three concurrent transactions: \textit{\textbf{T1:} RangeScan(40, 56)}, \textit{\textbf{T2:} Insert(42, ``Record Data'')}, and \textit{\textbf{T3:} LookUp(50)}. \figref{fig:illus1} shows the database's initial state: the buffer caches R2, R45, R55, and R99, while R50, R70, and R88 exist in the remote storage. T1 starts first, followed by T2 and T3. We walk through the execution in \figref{fig:phantomillustration}. \looseness=-1

\boldtext{Transaction T1.}
T1 reads range $[40, 56]$. \proposed first inserts two sentinel records R40 and R56 at the range boundaries (\figref{fig:illus2}, \textbf{Step~1}). It then acquires \textbf{shared} next-key locks (S NK-locks) on R40, R45, and R55, protecting the cached records and their subsequent gaps $(40, 45)$, $(45, 55)$, and $(55, 56)$ (\textbf{Step~2}). Without sentinels, it would have to lock the smallest cached span, i.e., $[2, 99)$, introducing unnecessary contention. After locking, T1 follows \texttt{DB.RangeScan()} (as in \algoref{alg:operations}) to fetch missing data from remote storage. Since only R50 exists remotely, \proposed admits R50 to the buffer and converts the sentinels into tombstones (\textbf{Step~3}). 

Cache admission can insert records into the buffer even when an NK-lock is held. In this example, T1 admits R50 into the buffer despite the NK-lock on R45, because transactional locks only block user-initiated inserts into $[45, 55)$, not cache admissions, which change a record’s physical location, not the logical database contents. Since the predecessor of R50 (key 45) already holds an NK-lock, R50 is also NK-locked by the same transaction: the lock on R50 now protects the gap $[50, 55)$, and the lock on R45 implicitly shrinks to protect $[45, 50)$. These locks are held until commit or abort.

\boldtext{Transaction T2.} A concurrent T2 inserts a new record R42. For clarity, \figref{fig:pp_q2} shows T2 executing when T1 reaches step 3. It first checks whether it can acquire \textbf{exclusive} next-key locks (X NK-locks) on the gap covering R42. In this case, it finds R40, the largest indexed key smaller than or equal to R42. However, a shared NK-lock is already held by T1, so T2 either waits for T1 to release the lock or self-aborts (\textbf{Step~1}). 

\boldtext{Transaction T3.} T3 retrieves the record with key 50. For clarity, \figref{fig:pp_q3} shows T3 executing when T1 reaches step 3. Different from T1, T3 directly acquires a regular \textbf{shared} lock on R50 instead of an NK-lock (\textbf{Step~1}), because point reads cause no phantom issues. As R50 is already admitted by T1, T3 finds R50 in the buffer and directly returns it to the user (\textbf{Step~2}).

Note that sentinels and tombstones are managed exactly like regular records; the rules to maintain gap bits and dirty bits are orthogonal to the record type. On commit, the range query resolves sentinels into either regular records or tombstones according to the query results. On abort, the sentinels can be either removed immediately or left to be reclaimed by normal eviction.

\section{Durability and Recovery}
\label{sec:recovery}

Although the storage service is durable, its durability covers only flushed data. \proposed's record buffer is a write-back cache: committed updates are flushed lazily and may exist solely in compute memory, where a compute failure would lose them. Recovery must therefore restore these committed-but-unflushed updates. To this end, \proposed employs a decoupled recovery mechanism, similar to prior work~\cite{txn-hbase, Levandoski2011DeuteronomyTS} that supports transactions over key-value stores. Because the compute layer manages \textit{logical} records and is agnostic to storage layout, the redo log (WAL) tracks only logical record updates; structural changes to the in-memory tree (e.g., node splits, merges, and deletions) are not logged. On commit, the compute layer appends the transaction's updates to the WAL with the commit LSN, then applies them to the record buffer before releasing locks. The WAL resides in durable, highly available remote storage.

\boldtext{Decoupled Recovery.} In \proposed, compute nodes are decoupled from the remote storage service. We assume the storage service is durable and recovers itself upon failure. Accordingly, we focus on the compute layer and recover compute nodes only after the storage is restored. After a failure, a compute node starts with an empty record buffer and replays the WAL from the latest checkpoint LSN to the log tail, reconstructing the tree-structured buffer from logged record updates. The recovered buffer may differ from the pre-failure state because clean cached records are lost and lazily repopulated as new queries arrive. Updates from ongoing transactions that did not commit to the WAL are lost and treated as aborted.

\boldtext{Checkpointing.} To bound recovery time, \proposed performs periodic fuzzy checkpoints. At each checkpoint, we record the latest committed LSN $X$ and scan the buffer to asynchronously flush dirty records to remote storage. Since checkpointing proceeds concurrently with transaction processing, flushed data may include updates newer than $X$. This is safe because logical redo is idempotent: on recovery, the WAL is replayed from $X$ to the log tail to reconstruct the buffer. Any updates in storage already reflecting LSNs $> X$ are deterministically overwritten with identical values, ensuring the reconstructed state matches the log tail.

\section{Deployment in Cloud Databases}
\label{sec:deployment}

Our presentation so far assumes a single database instance. Production cloud databases distribute data across multiple compute nodes for scalability and have diverse storage-side designs. We show how to integrate \proposed into real cloud databases.  

\subsection{Scalable Compute Layer}
\label{ssec:partitioned_writer}

Classic storage-disaggregated databases such as Aurora~\cite{verbitski2017amazon}, PolarDB~\cite{PolarDB}, and Neon~\cite{neon} use a single-writer~\cite{ziegler2023scalable} architecture, where one primary node processes updates and read-only nodes serve reads. However, scalable OLTP systems increasingly adopt the \textit{partitioned-writer} architecture~\cite{ziegler2023scalable}, where data is partitioned across compute nodes, each with exclusive write ownership of its partitions. Aurora Limitless~\cite{aurora_limitless}, PolarDB-X~\cite{polardbx}, TiDB~\cite{tidb}, and CockroachDB~\cite{cockroachdb} follow this architecture. \proposed assumes a \emph{single writer per partition}: each partition is exclusively accessed by its owner compute node. The partitioned-writer architecture is thus a natural deployment target: (1)~\textit{No cross-node cache coherence.} Each compute node exclusively owns its partition, so its record buffer is authoritative for all cached data. No cross-node invalidation is needed, a major simplification over shared-storage designs. (2)~\textit{Independent per-node buffers.} Every node maintains its own tree-structured record buffer, with localized tombstones, gap bits, and sentinels. Thus, mechanisms like negative caching and phantom protection operate within a single node's partition, so compute scales by provisioning nodes for new partitions.

A transaction router, collocated with the compute layer or deployed as a dedicated tier, routes each query to the owning compute node. Within a compute node, conventional concurrency control (e.g., 2PL) ensures transactional consistency. Cross-partition transactions use standard two-phase commit (2PC)~\cite{cornus, Tapir}: each node acquires locks and buffers updates in its record buffer; the coordinator ensures atomic commitment. Phantom protection composes naturally with 2PC: a cross-partition range query splits at partition boundaries, and each participant inserts boundary sentinels and takes next-key locks over its sub-range, as in the single-node protocol (\secref{sec:phantomprotection}); the coordinator's atomic commit ties participants together. No cross-node coordination of tombstones, gap bits, or sentinels is needed, as each node's metadata covers only its partition.

Partition migration is handled by the host system's ownership-transfer mechanism, not by \proposed: the old owner flushes dirty records before releasing ownership, and the new owner starts cold and repopulates lazily, as in recovery (\secref{sec:recovery}). Live migration has established solutions~\cite{albatross, remus, squall, marlin}, and \proposed composes with them.

\subsection{Storage Layer Integration}
\label{ssec:storage_integration}

\proposed requires the storage to satisfy two conditions:
(1)~support the four record-based atomic APIs of~\tabref{tab:data_store_api}, and
(2)~provide internal indexing so index traversal can be pushed down. Conversely, stores lacking native ordered range scans (e.g., Redis~\cite{redis}, DynamoDB~\cite{dynamodb}) cannot host \proposed directly. We show how cloud databases and general-purpose storage services meet these requirements. \looseness=-1

\boldtext{Record-Oriented Cloud Databases.} As discussed in~\secref{ssec:remote-index}, many disaggregated cloud databases \cite{Levandoski2011DeuteronomyTS, tidb_serverless, aurora_dsql, yugabytedb, zhou2021foundationdb} unbundle the database kernel into transaction and data components. The data component usually provides record-oriented interfaces and internal indexing that directly satisfy \proposed's requirements.
For example, Deuteronomy exposes atomic
\textit{Read}, \textit{Update}, \textit{Insert}, and \textit{Delete}
operations, and uses the Bw-tree~\cite{Levandoski2013bwtree} for internal indexing.
TiDB's storage layer, TiKV, provides gRPC
APIs fulfilling all four \proposed interfaces, with an internal LSM-tree index.

\boldtext{Page-Oriented Cloud Databases.} Some cloud databases (e.g., Aurora~\cite{verbitski2017amazon} and PolarDB~\cite{PolarDB}) adopt a disaggregated architecture yet retain a page-based index: the storage layer exposes page-oriented interfaces, and the WAL consists of page-level redo records. Deploying \proposed on such systems is nontrivial and requires two major changes: (1) the storage layer must expose a record-level API atop its page servers, and (2) the WAL must shift to logical record-level operations, which requires adapting log replay and recovery.

\boldtext{General-Purpose Cloud Storage.}
Widely used cloud storage services meet \proposed's conditions natively.
\textit{\textbf{Azure Table Storage}}~\cite{azuretablestorage} provides
REST APIs with entity-level operations: \textit{Get} maps to a point
lookup by \textit{PartitionKey} and \textit{RowKey}, \textit{GetRange}
to \textit{Query Entities} with filters, \textit{Upsert} to
\textit{Insert Or Replace Entity}, and \textit{Delete} to
\textit{Delete Entity}. Its \textit{PartitionKey}--\textit{RowKey} primary index satisfies index pushdown.
\textit{\textbf{Google Cloud Datastore}}~\cite{google_cloud_datastore}
exposes a NoSQL interface via RPC: \textit{Get} and
\textit{GetRange} map to \textit{Lookup}
(\textit{projects.lookup}) and \textit{RunQuery} with range filters, while \textit{Upsert} and \textit{Delete} use
the atomic \textit{Commit} (\textit{projects.commit}) API.
Built-in indexes on entity properties support index pushdown. \looseness=-1

\section{Evaluation}
\label{sec:eval}

We evaluate \proposed against baseline designs and find: (1)~\textbf{\textit{Positive queries:}} \proposed beats page-based indexes, especially under skew (\secref{ssec:point_query}). (2)~\textbf{\textit{Negative searches:}} \proposed beats hash-indexed record caches, and page-based designs at low-to-moderate negative-query ratios, especially under skew (\secref{ssec:negative_search}). (3)~\textbf{\textit{Phantom protection:}} \proposed's tighter lock spans beat storage-agnostic designs (e.g., \pcl), especially in large, sparse key domains (\secref{ssec:phantom_protection}).

\subsection{System Configurations}
For fair comparison, we built a storage-disaggregated DBMS testbed\footnote{\url{https://github.com/hippohwj/Nemo-pub}} and integrated all baselines into it. Following \secref{sec:design}, the testbed includes a transaction-processing engine with buffering and a disaggregated storage service. It can be extended to a distributed database via partitioning; this does not affect our conclusions, as all nodes use the same indexing and buffering.

We use Azure Cosmos DB~\cite{cosmosdb} as the remote data store, configured with a 16K request-units-per-second throughput cap~\cite{azure_cosmosdb_ru}. The database server runs on an Ubuntu 18.04 Azure VM in the same region (West US 2), with an Intel Xeon Platinum 8272CL CPU (8 cores $\times$ 2 HT) and 64~GB DRAM. The testbed implements standard optimizations, including early lock release~\cite{gawlick1985varieties,orleans-txn,graefe2013controlled} and group commit~\cite{dewitt1984implementation}. We evaluate three configurations by default:

\boldtext{\proposed.} The record buffer is an in-memory B+tree (fanout 125). Cosmos DB's key-value interface meets the storage-API requirements of~\secref{sec:design}, and its built-in indexing is the remote index.

\boldtext{\btree Baseline.} We organize the index in 8~KB pages, one per index node. Each internal node stores child page IDs (4-byte integers). A leaf locates records via a page-ID/offset pair (each 4 bytes). Each node has 125 children. For scans, we link adjacent leaves bidirectionally. We use a clock-managed page buffer. Our \btree largely follows PostgreSQL~\cite{postgres}, except that disk I/O becomes network requests. We implement two variants: (1) \textit{\textbf{\btreeretain}} (Retain-Index) keeps index pages in memory and evicts data pages first; (2) \textit{\textbf{\btreeevict}} (Evict-Index) evicts index and data pages uniformly. \looseness=-1

\boldtext{\twotree~\cite{zhou2023two-tree} Baseline.} We implement the top tree in record granularity, while the bottom tree is a conventional page-based \btree. All index and data pages initially reside in remote storage and are fetched on demand. Each tree maintains its own buffer pool and evicts when full. We split the total buffer budget: 1~GB is reserved for the bottom tree to keep cold-index pages resident, and the remainder goes to the top tree. If the total budget is below 1~GB, we split capacity evenly. In the top tree, buffer saturation migrates records to the bottom tree; in the bottom tree, it triggers page eviction to remote storage. We evaluate \twotreeretain and \twotreeevict variants.

\subsection{Benchmarks}
\label{ssec:bench}
We evaluate with YCSB and TPC-C: (1) \textbf{YCSB.}
We use YCSB~\cite{cooper2010benchmarking} with a 10~GB dataset by default. Each record is about 1~KB. Unless stated otherwise, each transaction issues 16 operations, following common practice~\cite{ccbench,dbexp,yu2014staring}; we use read-only and read-write (80\% reads, 20\% updates) mixes with Zipfian skew $\theta=0$--$0.99$. Phantom-protection experiments follow YCSB-E with concurrent range scans and inserts (\secref{ssec:phantom_protection}). The default range length is 8, varied in some experiments. (2) \textbf{TPC-C.} 
We run TPC-C~\cite{tpcc} with 34 warehouses, nine tables plus the customer last-name index, variable-length records (16 bytes to hundreds), and the standard transaction mix. Transactions are multi-statement with inserts, range scans, and updates. Per data point, we warm the cache with the measurement workload until the buffer fills, then measure for 60 seconds. We use 16 workers by default. \looseness=-1

\subsection{Positive Search}\label{ssec:point_query}

\begin{figure}
  \vspace{-0.1in}
  \centering
\begin{subfigure}{0.35\linewidth}%
\centering
  \includegraphics[width=\linewidth]{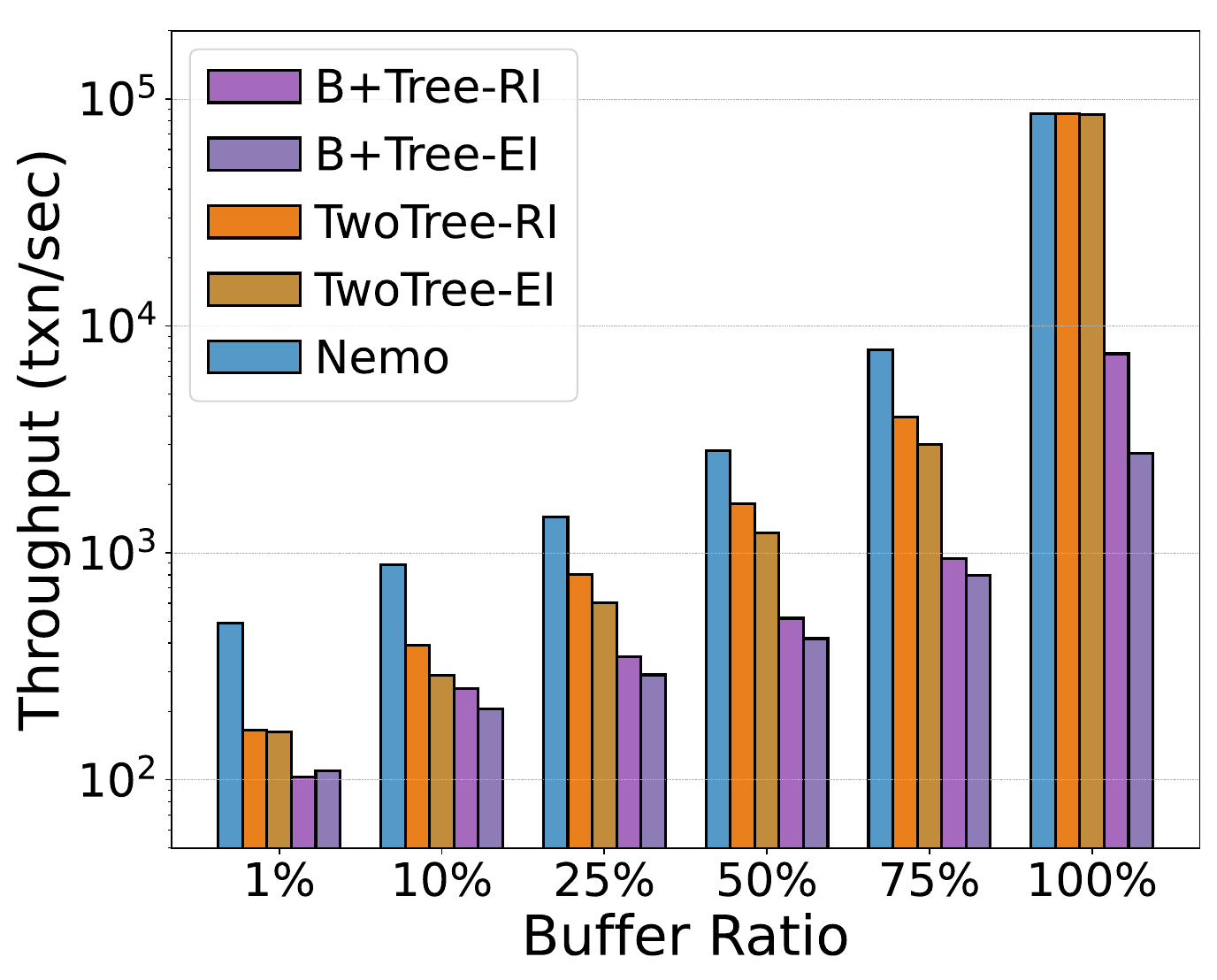}
  \caption{Throughput}
  \label{fig:exp1_f1_a}
\end{subfigure}\quad%
\begin{subfigure}{0.35\linewidth}%
\centering
  \includegraphics[width=\linewidth]{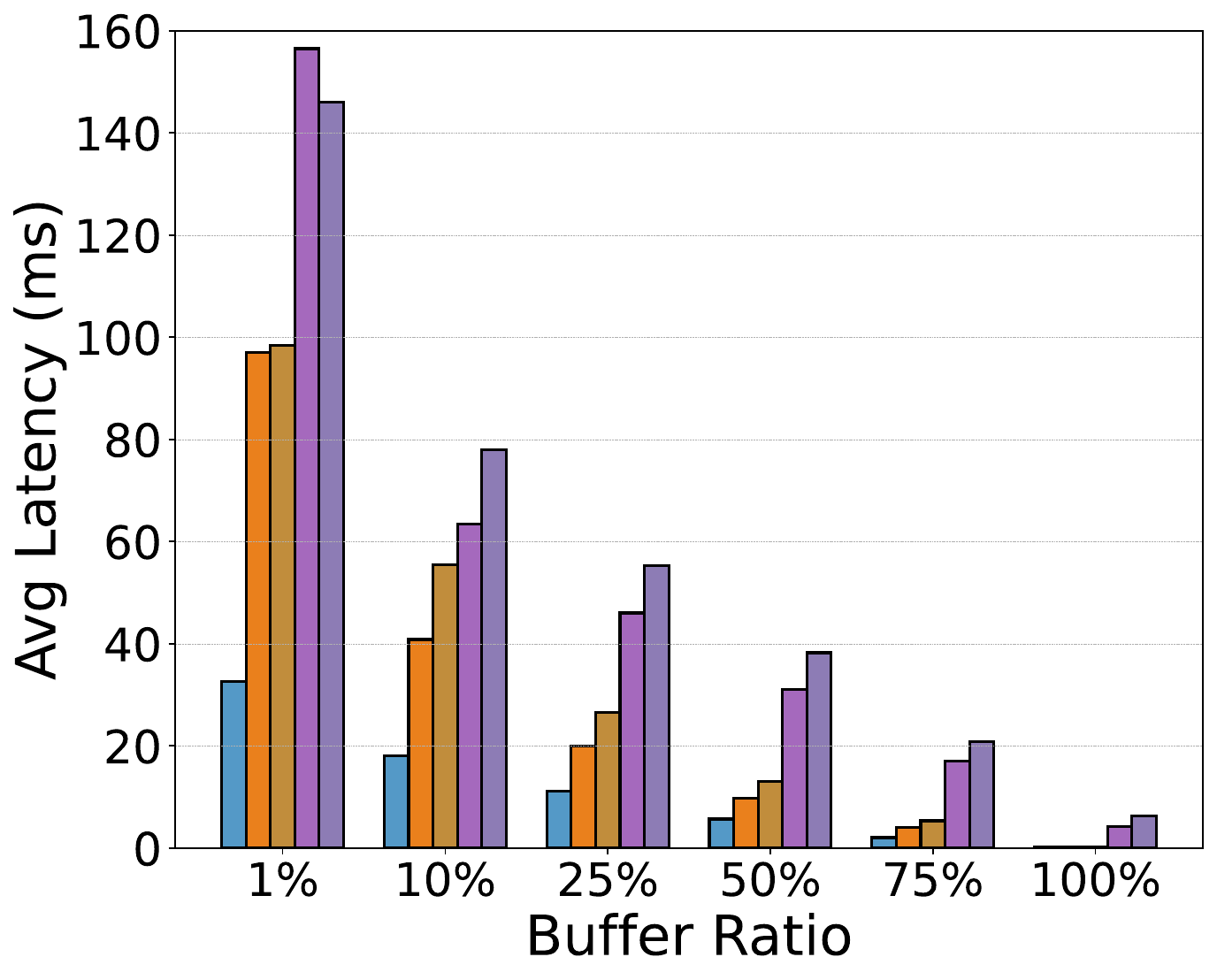}
  \caption{Average Latency}
  \label{fig:exp1_f1_b}
\end{subfigure}

\begin{subfigure}[t]{0.35\linewidth}%
\centering
  \includegraphics[width=\linewidth]{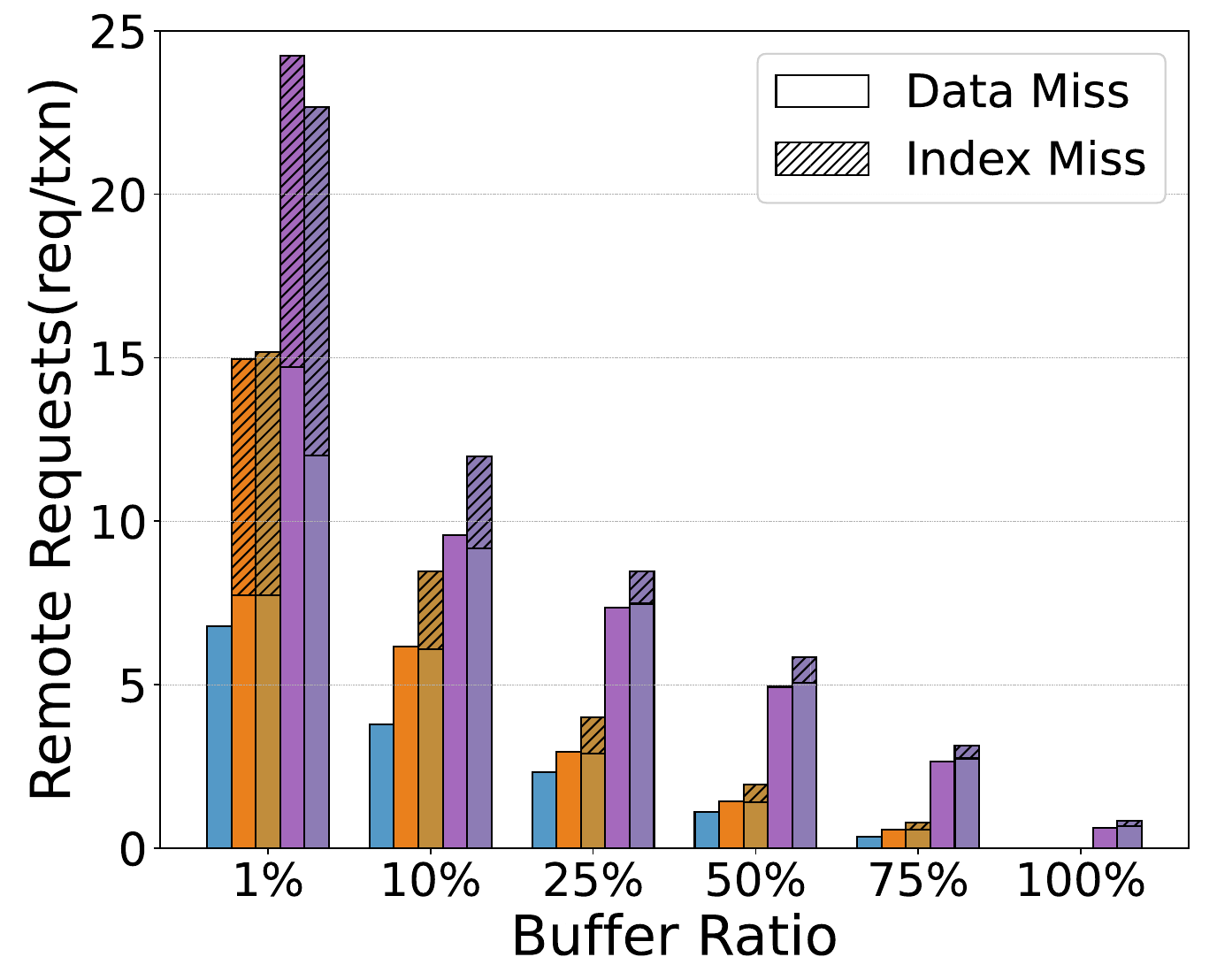}
  \caption{Remote Requests Number}
  \label{fig:exp1_f1_d}
\end{subfigure}\quad%
\begin{subfigure}[t]{0.35\linewidth}%
\centering
  \includegraphics[width=\linewidth]{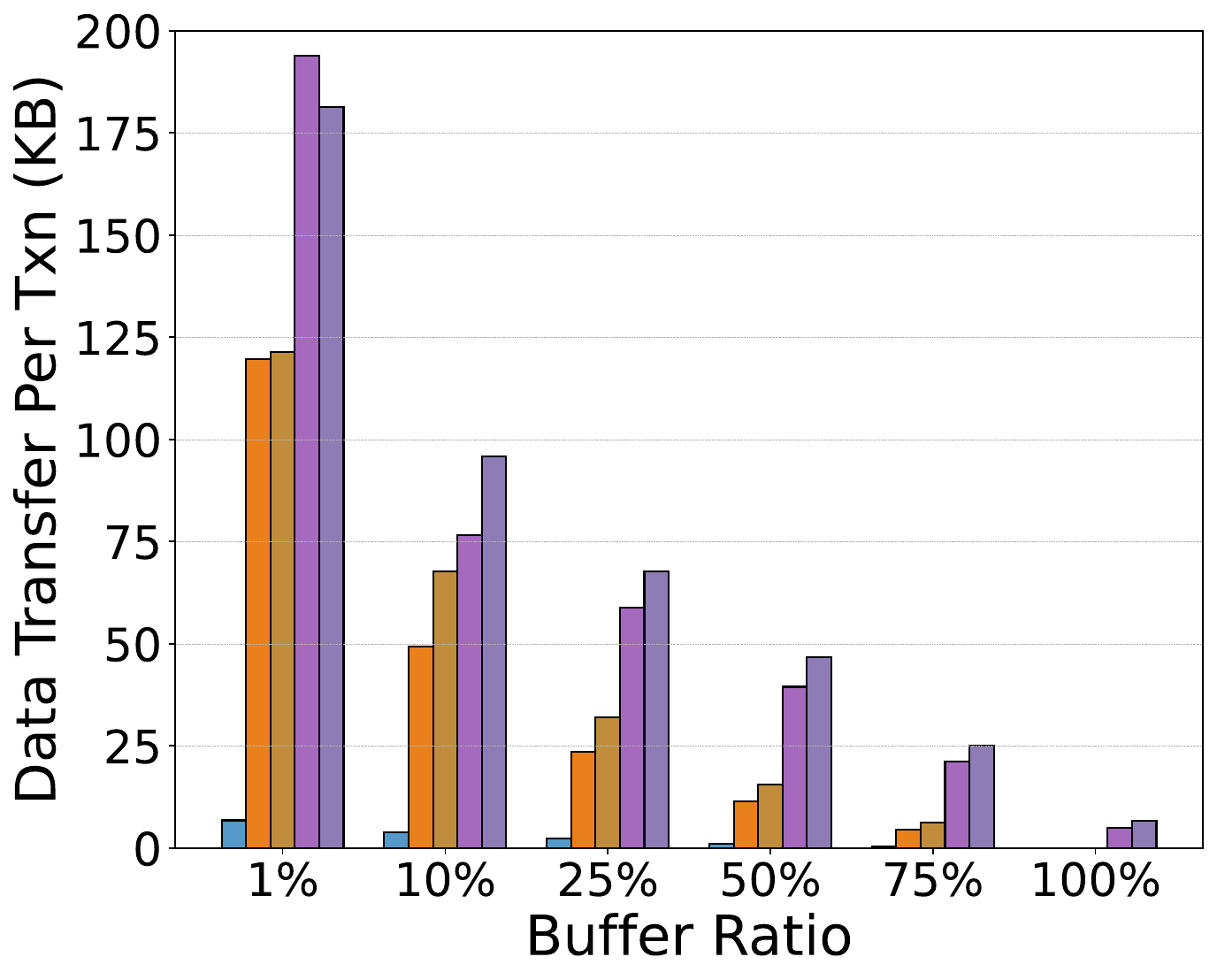}

\caption{Remote Data Transfer}
  \label{fig:exp1_f1_e}
\end{subfigure}%
\vspace{-0.08in}
\caption{Read-Only Performance Varying Buffer Size}
\label{fig:exp1_f1}
  \Description{Four grouped bar charts over buffer ratios of 1, 10, 25, 50, 75, and 100 percent, comparing B+Tree-RI, B+Tree-EI, TwoTree-RI, TwoTree-EI, and Nemo. Throughput rises and latency falls for every design as the buffer grows; the B+Tree variants are generally the slowest, and Nemo reaches up to 2.97 times the throughput of 2-Tree and 11.42 times that of B+Tree. The lower charts split remote requests per transaction into data misses and index misses and show kilobytes transferred per transaction; the page-based designs generally issue more requests and move far more data than Nemo.}
\end{figure}

\begin{figure}[t]
	\vspace{-0.1in}
        \centering
    \begin{subfigure}[t]{0.35\linewidth}%
    	\centering
        \includegraphics[width=\linewidth]{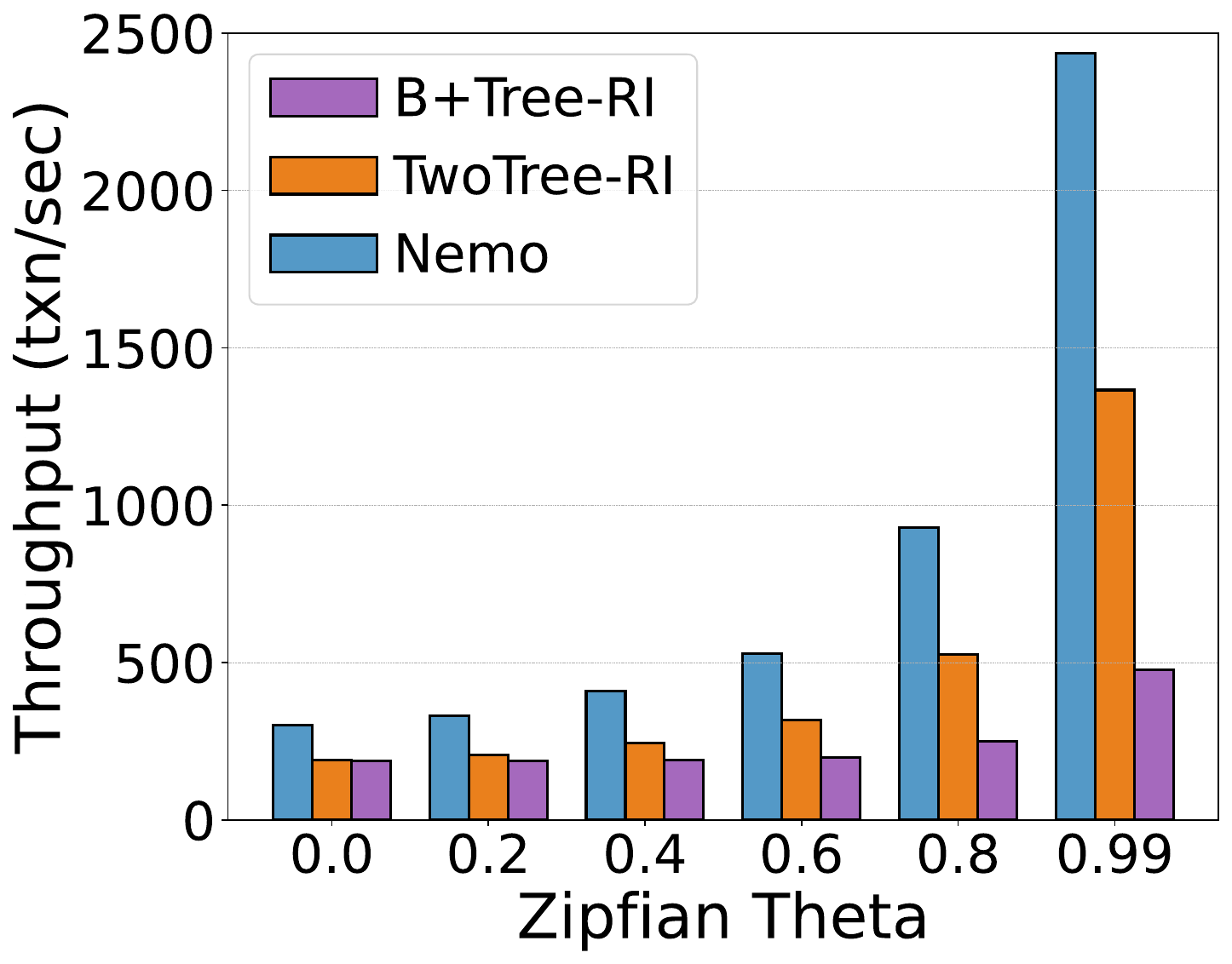}
        \caption{Read-Only Throughput}
        \label{fig:exp2_f1_a}
    \end{subfigure}\quad%
    \begin{subfigure}[t]{0.35\linewidth}%
    	\centering
        \includegraphics[width=\linewidth]{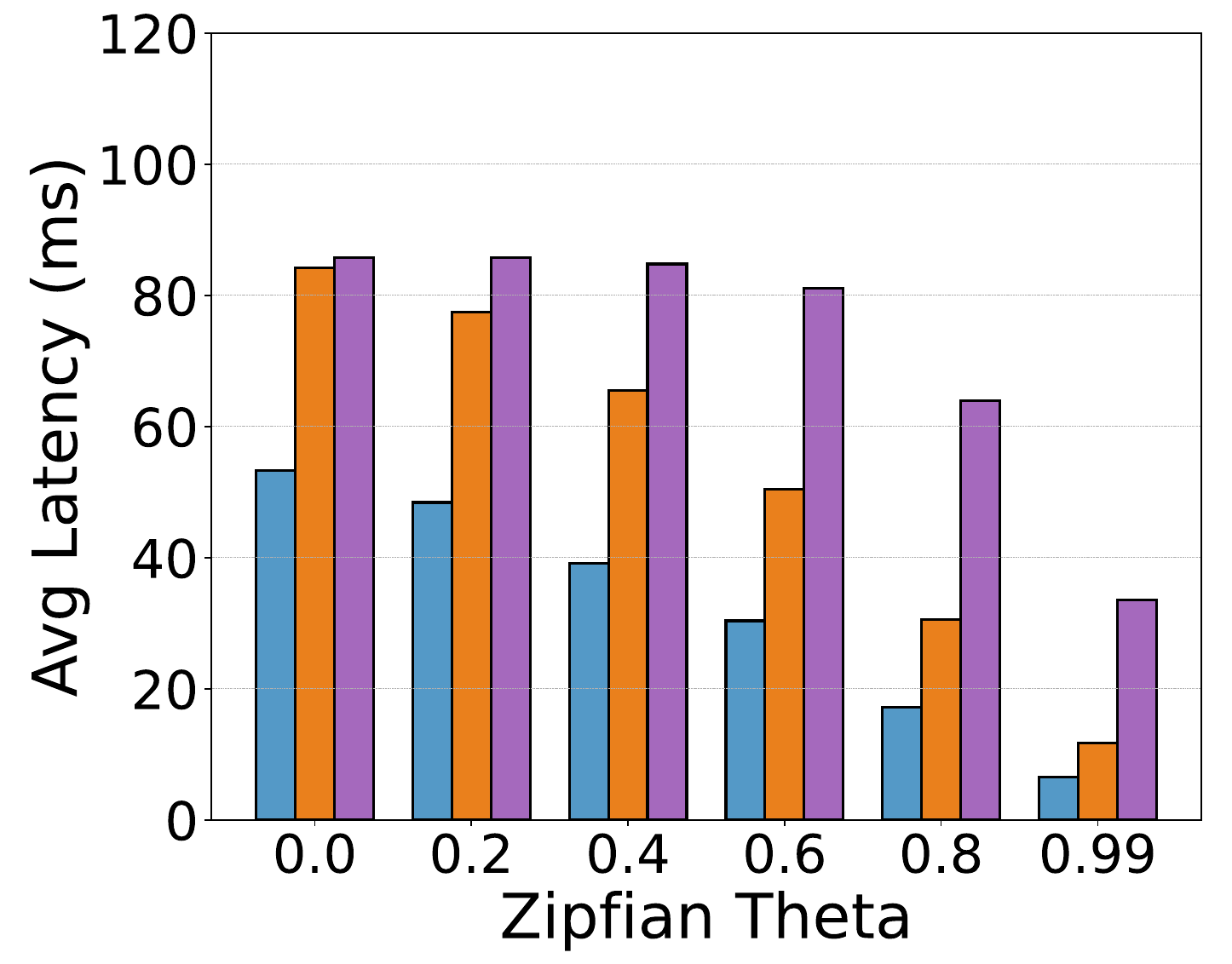}

        \caption{Read-Only Average Latency}
        \label{fig:exp2_f1_b}
    \end{subfigure}

    \begin{subfigure}[t]{0.35\linewidth}%
    	\centering
        \includegraphics[width=\linewidth]{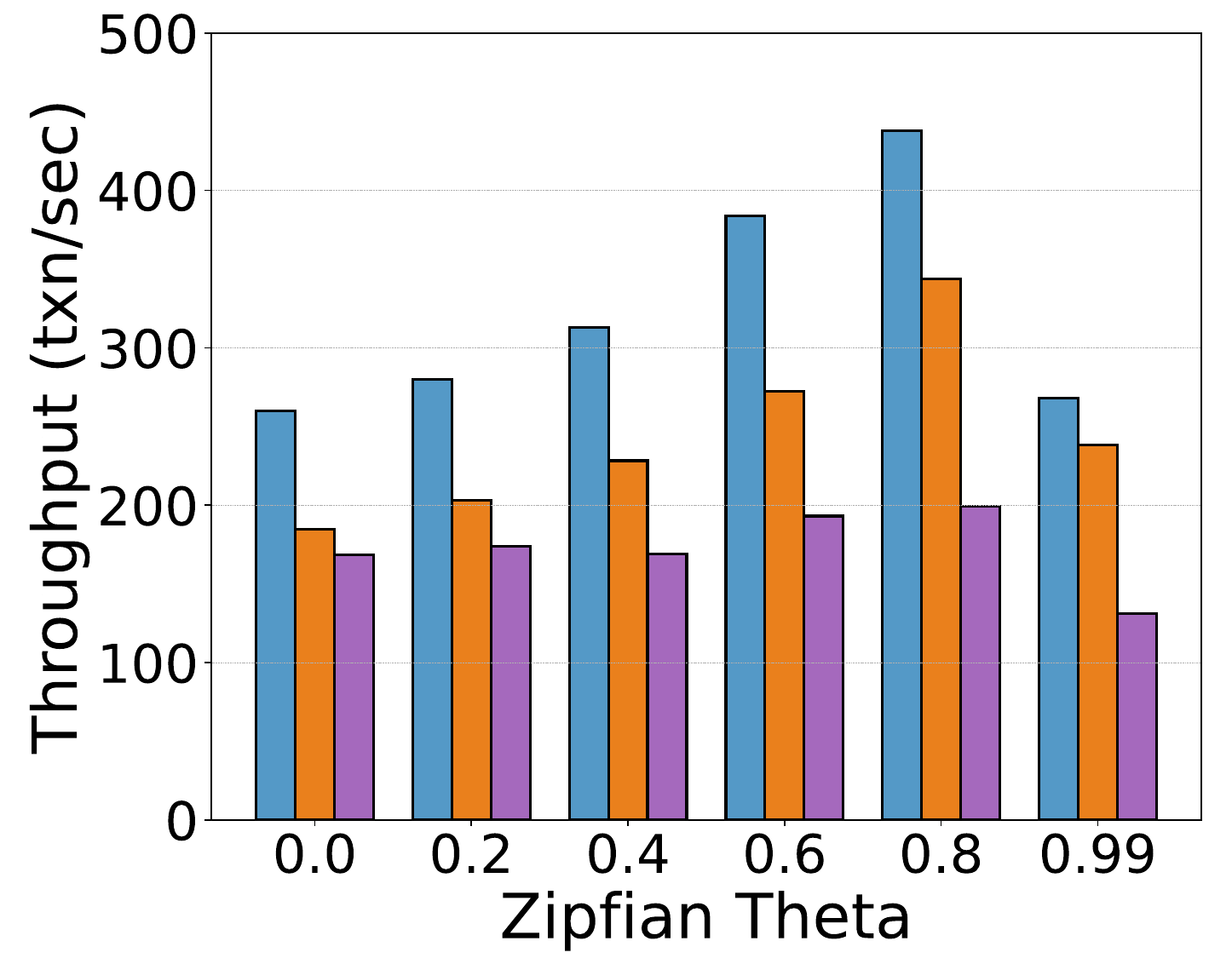}
      
     \caption{Read-Write Throughput}
        \label{fig:exp2_f1_c}
    \end{subfigure}\quad%
    \begin{subfigure}[t]{0.35\linewidth}%
    	\centering
            \includegraphics[width=\linewidth]{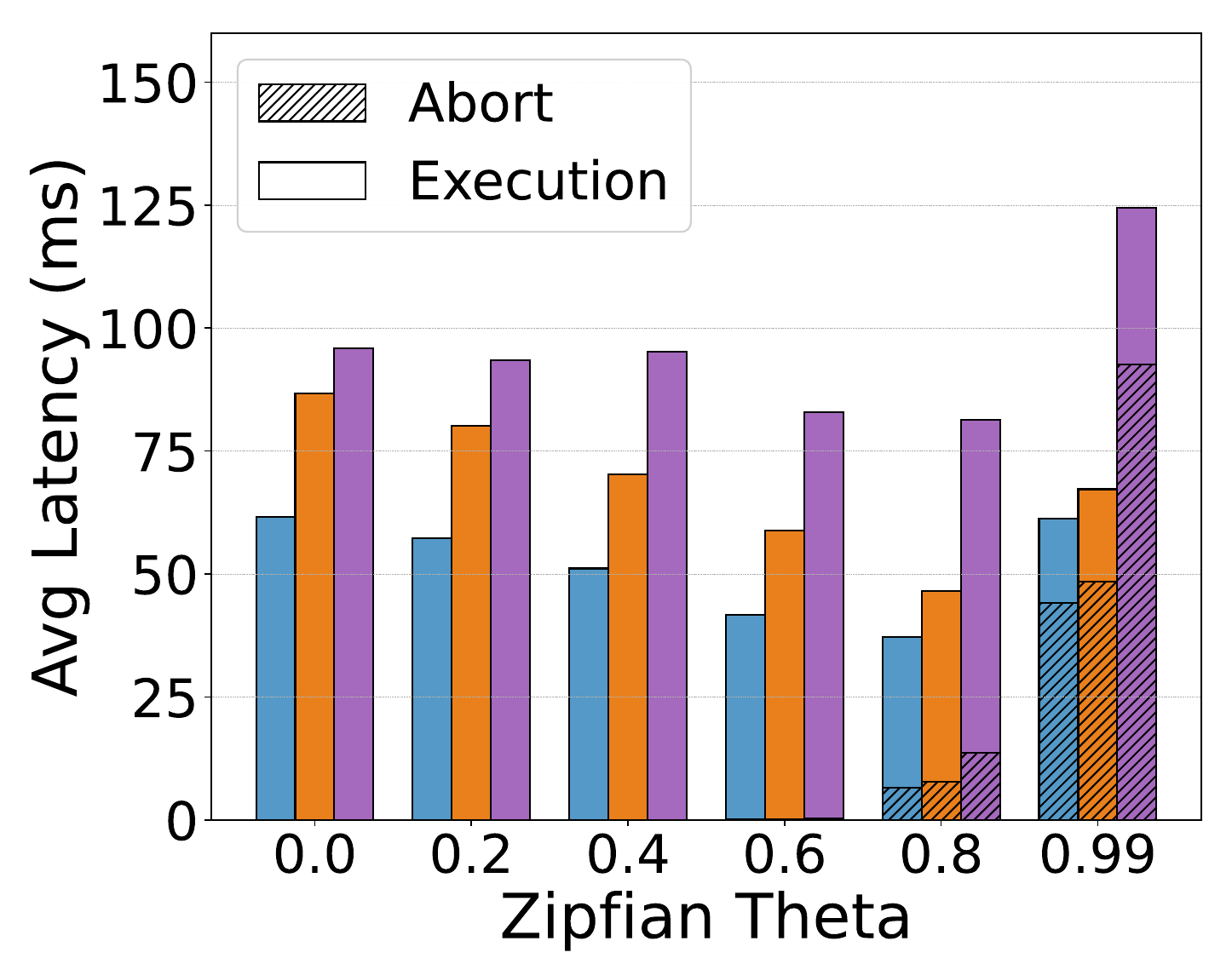}
     \caption{Read-Write Average Latency}
        \label{fig:exp2_f1_d}
    \end{subfigure}%

    \vspace{-0.08in}
    \caption{Performance Varying Skewness}
    \label{fig:exp2_f1}
  \Description{Four grouped bar charts over Zipfian theta from 0 to 0.99 at a 25 percent buffer ratio, comparing B+Tree-RI, TwoTree-RI, and Nemo. In the read-only workload, throughput rises with skew for all designs and Nemo leads, with speedups growing from 1.60 to 5.09 times over B+Tree and from 1.57 to 1.78 times over 2-Tree. In the read-write workload, throughput first rises and then drops at high skew, and latency bars split into abort and execution time show aborts dominating at theta 0.99.}
\end{figure}

This section evaluates \proposed and index baselines on positive point queries. Each transaction issues 16 requests, each a single-record read or update. We model skew with a Zipfian distribution parameterized by $\theta$, where $\theta=0$ is uniform and larger $\theta$ implies higher skew. We run two experiments, varying buffer size (\figref{fig:exp1_f1} and \figref{fig:exp1_f2}) and query skew (\figref{fig:exp2_f1}). For each, we evaluate read-only (100\% reads) and read-write (80\% reads, 20\% updates) workloads.

\subsubsection{Buffer Size}
We use a skewed query distribution ($\theta=0.9$) and vary the \textit{buffer size}, i.e., compute-node memory caching data and indexes. We study how buffer ratio ($\frac{\textit{Buffer Size}}{\textit{Data Size}}$) affects each design. For \btree and \twotree, the buffer ratio must exceed 100\% to fit both the data and the index in memory. \looseness=-1

\boldtext{Read-Only.} We first focus on the read-only results in \figref{fig:exp1_f1}. For all designs, throughput increases and latency decreases as buffer ratio grows. Overall, \btree performs the worst, whether it retains index nodes (\btreeretain) or not (\btreeevict), due to three reasons:

(1) \textit{Low Cache Efficiency.} \btree must cache a full page even if only one record is hot, wasting cache space and triggering more remote requests. As shown in \figref{fig:exp1_f1_d}, \btree issues many more remote requests per transaction than \proposed or \twotree.

(2) \textit{Network Traffic Amplification.} \btree fetches a full 8 KB page even for one 1 KB record, increasing per-request latency. \btree and \twotree average 6.58 ms per request, 40\% longer than \proposed (4.69 ms). \figref{fig:exp1_f1_d} and \figref{fig:exp1_f1_e} confirm that \btree transfers far more data per transaction.

(3) \textit{Serial Network Round-Trips.} \btree traverses the index at the compute node, causing serial network round-trips per index traversal. \btreeretain mitigates this by caching index pages, but this costs buffer space that could hold hot data, and it still fetches index pages when memory is tight. At 1\% buffer ratio, \btreeretain incurs about 9.5 index-page misses per transaction in \figref{fig:exp1_f1_d}, so at least one data miss triggers two or more sequential network round-trips.

\twotree outperforms \btree by separating hot and cold data into two trees. However, \twotree stores the bottom tree in pages like \btree, consuming substantial buffer. \figref{fig:exp1_f1_d} shows \twotree generally issues more remote requests per transaction than \proposed, especially when buffer ratio is small. \twotree also suffers network-traffic amplification (\figref{fig:exp1_f1_d} and \figref{fig:exp1_f1_e}) because its disk-oriented design still moves data in pages. It can also incur multiple sequential trips for index traversal, especially when index pages do not fit or can be evicted (\twotreeevict), as shown in \figref{fig:exp1_f1_d}. \looseness=-1

\begin{figure}
  \vspace{-0.1in}
  \centering
\begin{subfigure}[t]{0.35\linewidth}%
\centering
  \includegraphics[width=\linewidth]{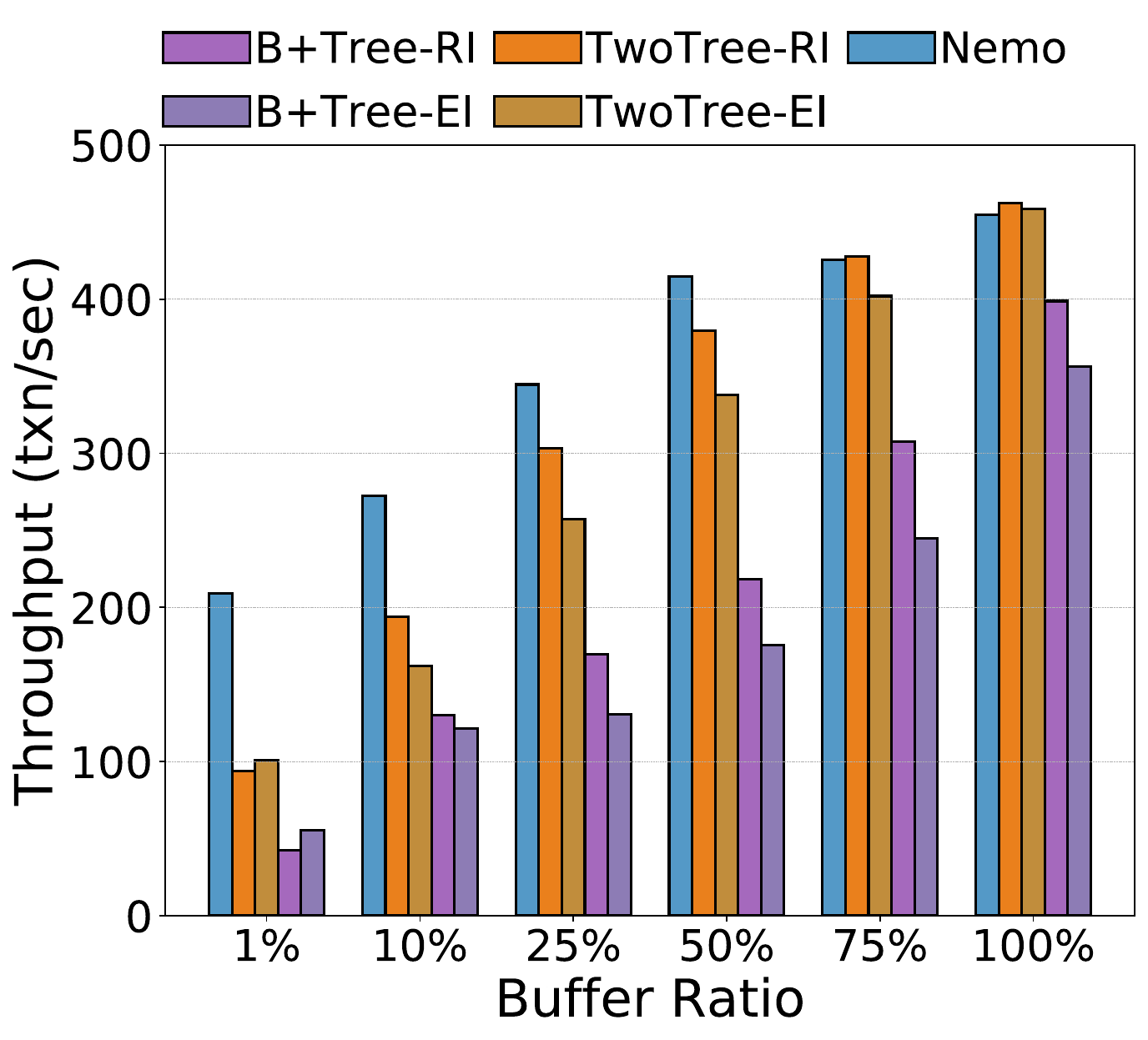}        
  \caption{Throughput}
  \label{fig:exp1_f2_a}
\end{subfigure}\quad%
\begin{subfigure}[t]{0.35\linewidth}%
\centering
  \includegraphics[width=\linewidth]{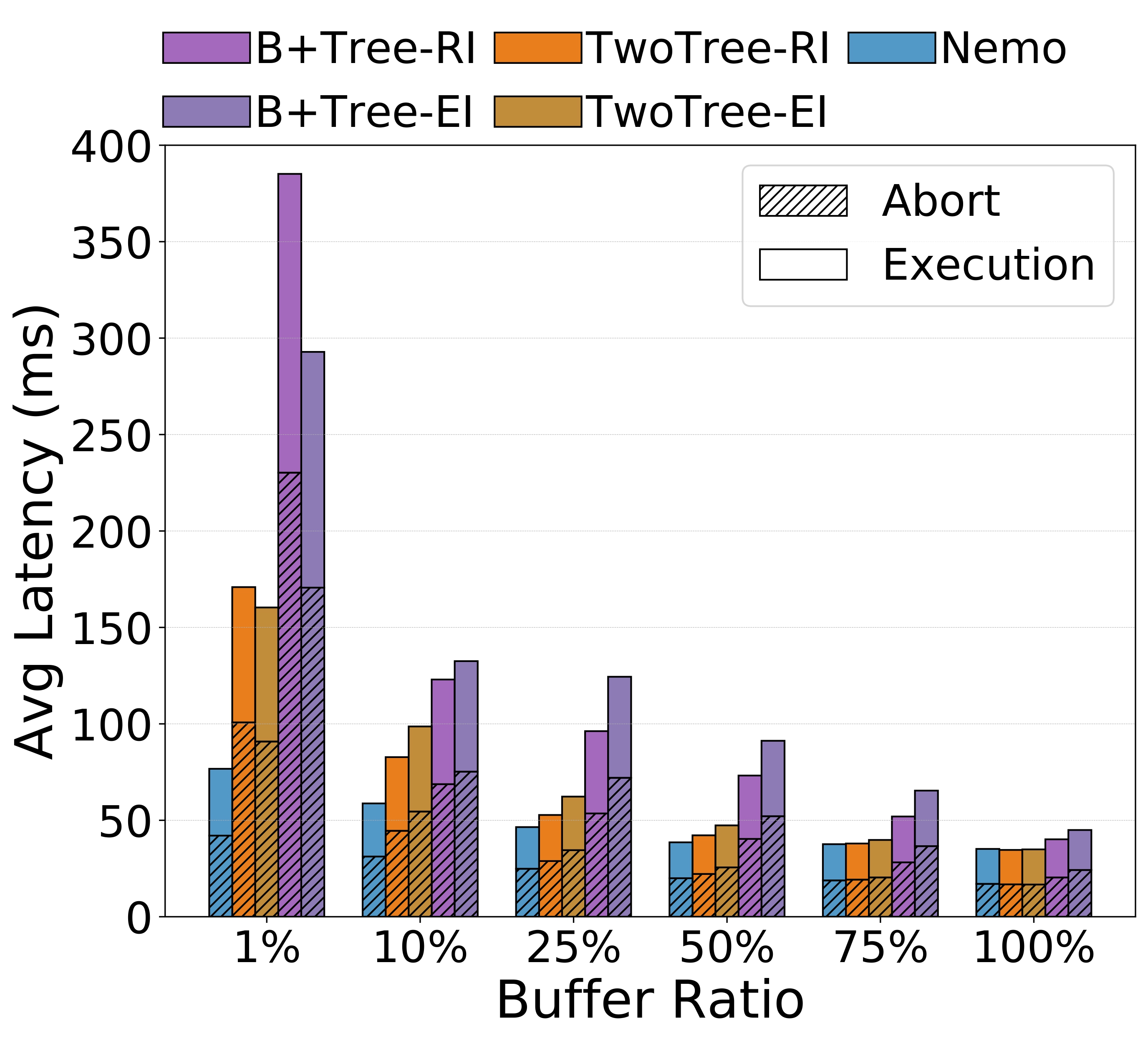}        
  \caption{Average Latency}
  \label{fig:exp1_f2_b}
\end{subfigure}%

\vspace{-0.08in}
\caption{Read-Write Performance Varying Buffer Size}
\label{fig:exp1_f2}
  \Description{Two grouped bar charts over buffer ratios from 1 to 100 percent for the read-write workload, comparing B+Tree-RI, B+Tree-EI, TwoTree-RI, TwoTree-EI, and Nemo. Nemo has the highest throughput, up to 3.78 times that of B+Tree, and its lead narrows to near parity as the buffer grows. Latency bars are split into abort and execution time, and abort time dominates.}
\end{figure}

\proposed addresses all three issues, achieving up to 2.97$\times$ speedup over \twotree and 11.42$\times$ over \btree (at 100\% buffer ratio), where \proposed serves all queries locally while \btree still suffers misses from index pages and cold records.

\boldtext{Read-Write.} \figref{fig:exp1_f2} reports results for read-write workloads. The trend matches read-only: \proposed still outperforms the baselines, reaching up to 3.78$\times$ speedup. As buffer ratio increases, \proposed's speedup over \twotree shrinks from 2.07$\times$ to parity, and over \btree drops from 3.78$\times$ to near parity. With more data cached, higher hit rates diminish \proposed's cache and network efficiency advantages. Speedups are smaller than in read-only because contention and logging offset these benefits. The latency breakdown in \figref{fig:exp1_f2_b} shows abort time dominates total latency.

For both \btree and \twotree, \indexretain generally outperforms \indexevict unless the buffer is very small (<5\%). Retaining index pages reduces serial network trips for index traversal. However, when the buffer is smaller than the index, \indexretain still keeps cold index pages in cache rather than hot data, causing more remote requests than \indexevict.

\subsubsection{Skewness}

In \figref{fig:exp2_f1}, we fix the buffer ratio at 0.25 and change skew by varying Zipfian $\theta$ from 0 to 0.99. Since \indexretain outperforms \indexevict at 0.25, we only show \btreeretain and \twotreeretain here.

In the read-only scenario (\figref{fig:exp2_f1_a}, \figref{fig:exp2_f1_b}), throughput of all systems increases with skew as cache hit rates improve. \proposed's remote requests per transaction drop from 11.36 to 1.32 as $\theta$ increases. Its speedup over \btree and \twotree grows from 1.60$\times$ to 5.09$\times$ and from 1.57$\times$ to 1.78$\times$, respectively, due to better memory use and less traffic amplification. There is no gain from reducing index-traversal trips because the buffer can cache all index pages for \btreeretain and \twotreeretain.

\figref{fig:exp2_f1_c} and \ref{fig:exp2_f1_d} show that \proposed still performs best in the read-write scenario. For all systems, throughput first rises but later drops as skew grows. Higher skew improves cache hits but concentrates accesses, increasing conflicts; throughput declines when contention outweighs caching gains. As evident in \figref{fig:exp2_f1_d}, abort time dominates latency at $\theta=0.99$. Under high skew, \proposed{}’s speedup over other baselines narrows because its cache-efficiency advantage is offset by contention overhead. \looseness=-1

Overall, \proposed shows a maximum throughput gain of 11.42$\times$ over \btree at a setting where buffer ratio = 100\% and 2.97$\times$ over \twotree at another setting where buffer ratio = 1\% in \figref{fig:exp1_f1_a}.

\subsection{Sensitivity and Ablation Studies}
\label{ssec:sensitivity}
We examine whether the conclusions of \secref{ssec:point_query} hold beyond its default setting.

\subsubsection{Sensitivity to Data Scale} \label{sssec:large_dataset}

\begin{figure}
  \vspace{-0.1in}
  \centering
  \begin{subfigure}{0.35\linewidth}
    \centering
    \includegraphics[width=\linewidth]{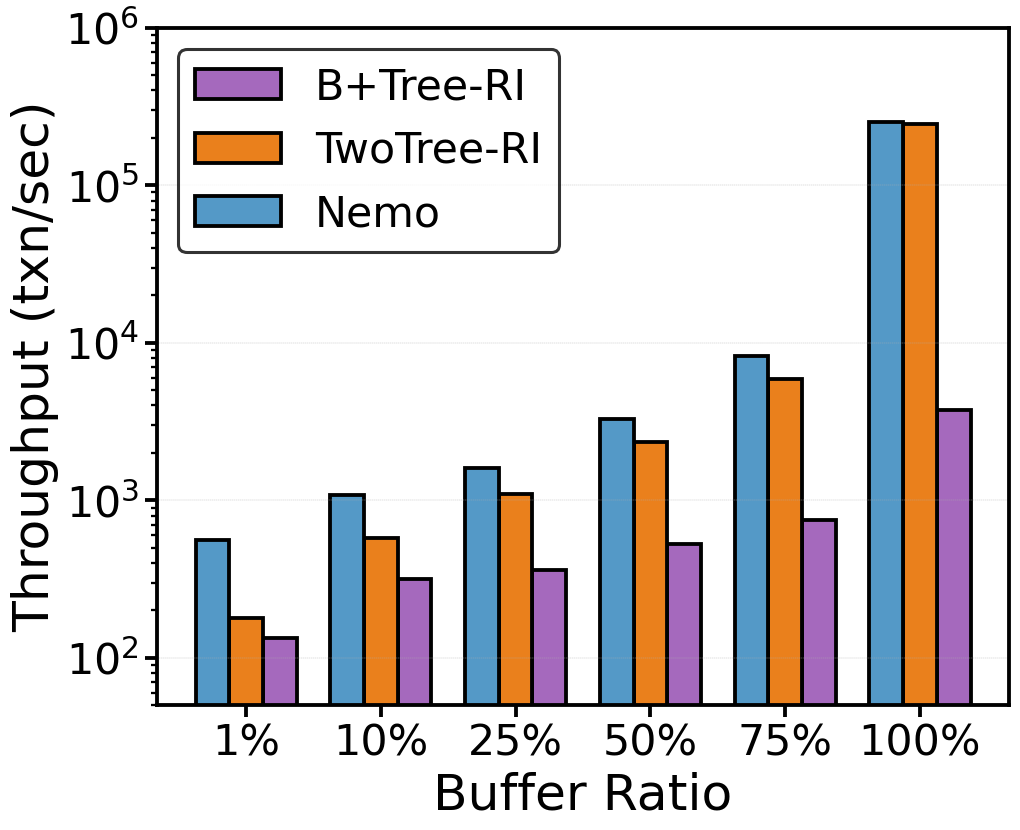}
     \caption{Read-Only Throughput}
    \label{fig:ycsb100g_a}
  \end{subfigure}\quad%
  \begin{subfigure}{0.35\linewidth}
    \centering
    \includegraphics[width=\linewidth]{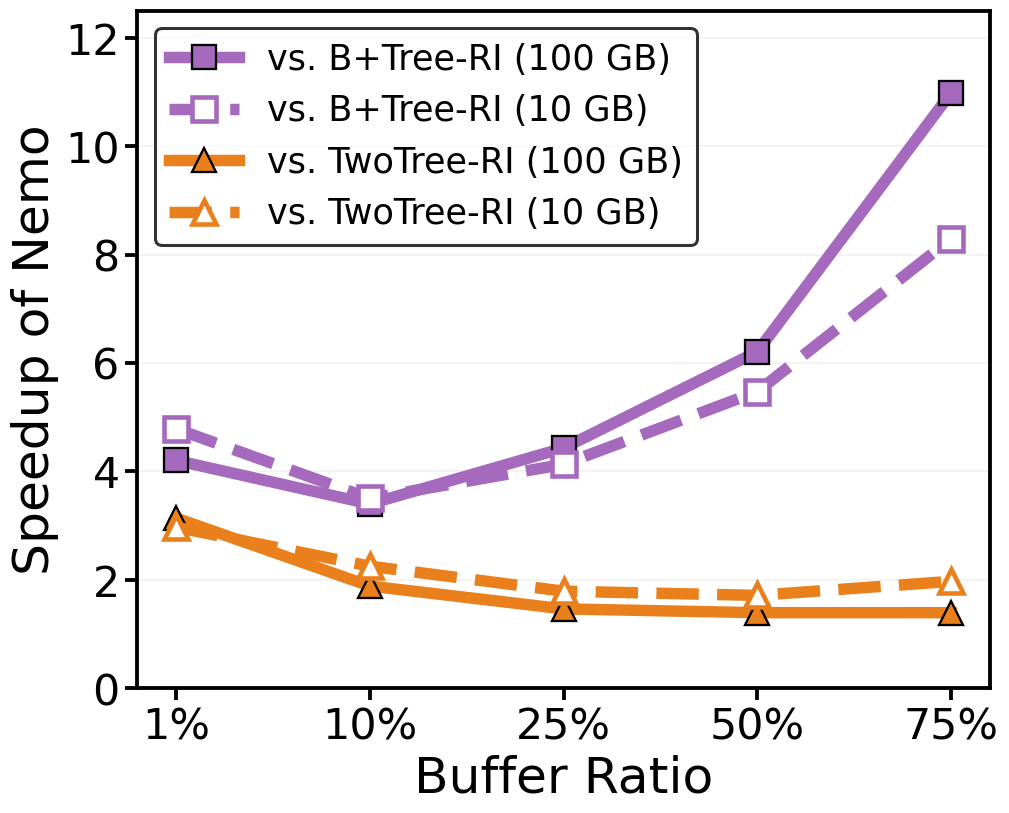}
     \caption{Speedup: 10~GB vs.\ 100~GB}
    \label{fig:ycsb100g_b}
  \end{subfigure}
  \vspace{-0.08in}
  \caption{Sensitivity to Data Scale (100~GB YCSB)}
  \label{fig:ycsb100g}
  \Description{Left: grouped throughput bars on a log scale over buffer ratios from 1 to 100 percent for B+Tree-RI, TwoTree-RI, and Nemo on the 100 GB dataset; Nemo leads at every ratio. Right: lines of Nemo's speedup over B+Tree-RI and over TwoTree-RI for the 10 GB and 100 GB datasets. The speedup over B+Tree-RI grows with the buffer ratio, the speedup over TwoTree-RI falls from about three times toward parity, and the 10 GB and 100 GB curves for each baseline nearly coincide.}
\end{figure}

This experiment tests whether our conclusions extend to larger datasets. We repeat the buffer-size sweep of \figref{fig:exp1_f1} ($\theta=0.9$) on a 100~GB YCSB dataset, using a larger VM (48 vCPUs, 192~GB DRAM) so every buffer ratio fits in memory. \figref{fig:ycsb100g_a} reproduces the prior trends: \proposed leads at every buffer ratio, its margin over \btreeretain rising past an order of magnitude with the buffer and its margin over \twotreeretain shrinking from about 3$\times$ to near parity; at 100\%, \proposed and \twotreeretain serve all requests locally while \btreeretain still cannot fit its data and index pages. \figref{fig:ycsb100g_b} plots \proposed's speedup over each baseline against the buffer ratio, for both the 10~GB and 100~GB datasets; for each baseline the two curves nearly coincide. \proposed's advantage is thus governed by the buffer ratio, not by the absolute data size.

\subsubsection{Sensitivity to Workload} \label{sssec:tpcc}

\begin{figure}
  \vspace{-0.1in}
  \centering
  \begin{subfigure}{0.35\linewidth}
    \centering
    \includegraphics[width=\linewidth]{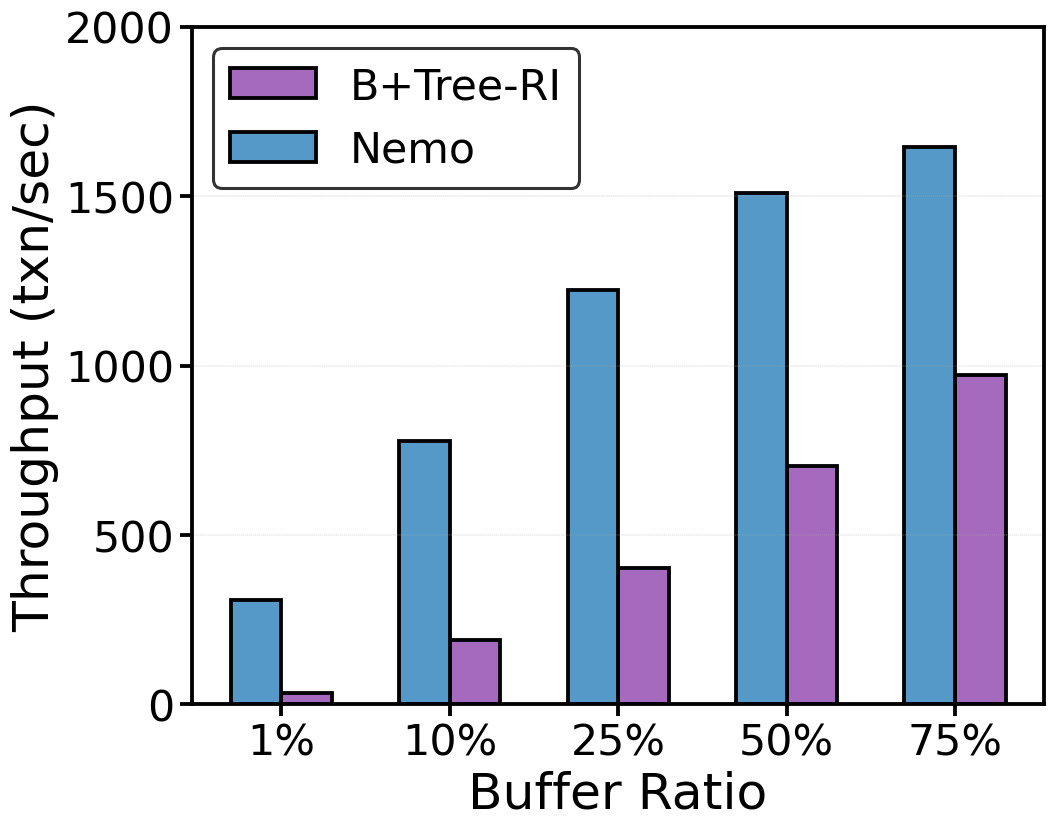}
     \caption{Throughput}
    \label{fig:tpcc_a}
  \end{subfigure}\quad%
  \begin{subfigure}{0.35\linewidth}
    \centering
    \includegraphics[width=\linewidth]{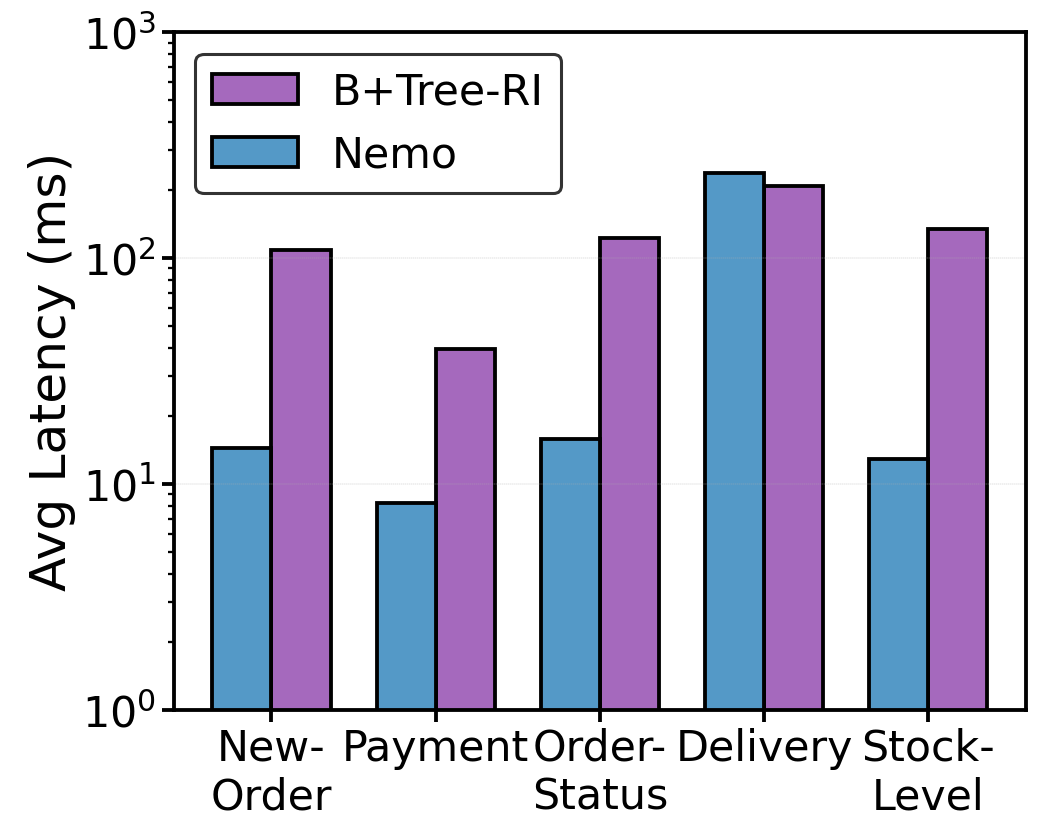}
     \caption{Transaction Latency (10\% Buffer)}
    \label{fig:tpcc_b}
  \end{subfigure}
  \vspace{-0.08in}
  \caption{TPC-C Performance Varying Buffer Size}
  \label{fig:tpcc}
  \Description{Left: throughput bars of B+Tree-RI and Nemo on TPC-C for buffer ratios from 1 to 75 percent; Nemo is faster at every ratio, by close to ten times with a small buffer and 1.7 times at 75 percent. Right: average latency of each transaction type on a log scale; Nemo is 5 to 8 times faster on New-Order, Payment, and Order-Status, about ten times faster on Stock-Level, and on par on Delivery.}
\end{figure}

This experiment evaluates \proposed on a realistic OLTP workload. \figref{fig:tpcc_a} compares \proposed and \btreeretain on TPC-C (\secref{ssec:bench}). \proposed is faster at every buffer ratio, by close to 10$\times$ when memory is tight and still by 1.7$\times$ when the buffer grows to 75\% of the dataset.  TPC-C's smaller, variable-size records amplify \proposed's advantages relative to the YCSB setting with fixed 1~KB records. With a tight buffer, index pages take up most of \btreeretain's frames, so it needs over three times as many remote reads per transaction as \proposed; with a large buffer both miss about equally often, and the gap is the cost of fetching an 8~KB page instead of a 100-byte record. \looseness=-1

\figref{fig:tpcc_b} shows the average latency of each transaction type at a 10\% buffer. \proposed is 5--8$\times$ faster on New-Order, Payment and Order-Status, which are built from point lookups and inserts. Stock-Level scans recently inserted order-lines, and \proposed is about ten times faster because recently inserted records stay cached locally, while \btreeretain misses stock/order-line pages at this buffer size. Delivery, dominated by per-district range scans, is on par.

\subsubsection{Ablation: Remote Storage} \label{sssec:sim_storage}

\begin{figure}
  \vspace{-0.1in}
  \centering
  \begin{subfigure}{0.35\linewidth}
    \centering
    \includegraphics[width=\linewidth]{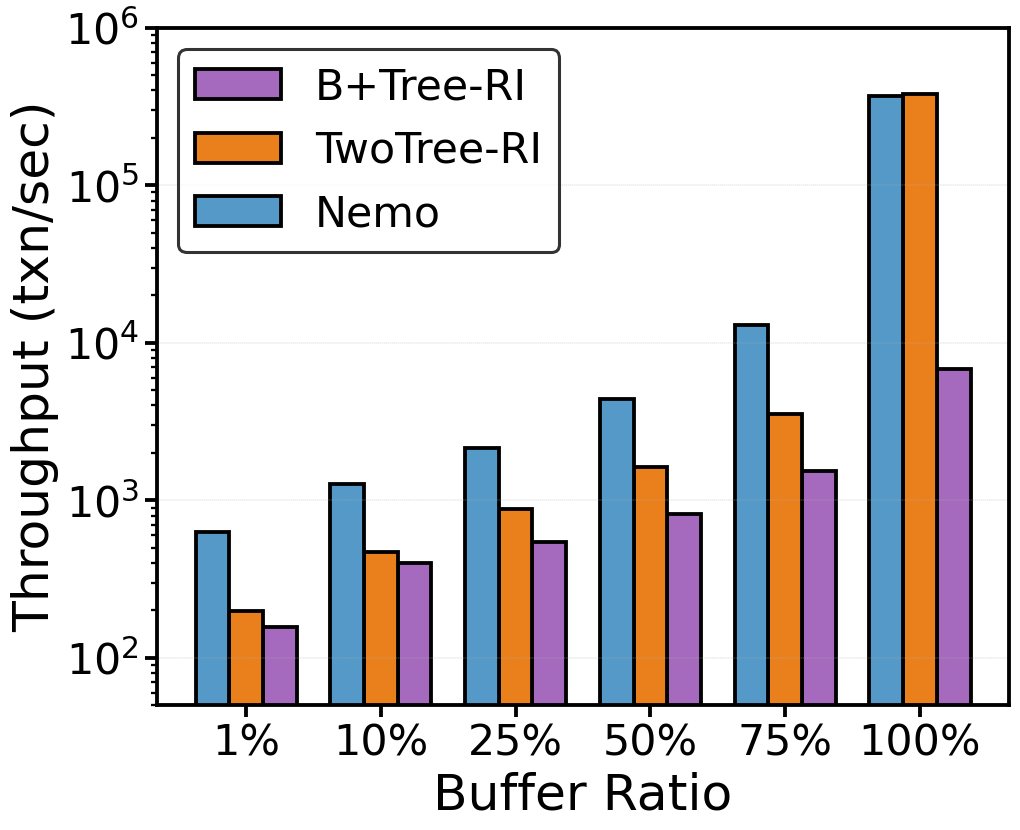}
     \caption{Read-Only Throughput}
    \label{fig:sim_store_a}
  \end{subfigure}\quad%
  \begin{subfigure}{0.35\linewidth}
    \centering
    \includegraphics[width=\linewidth]{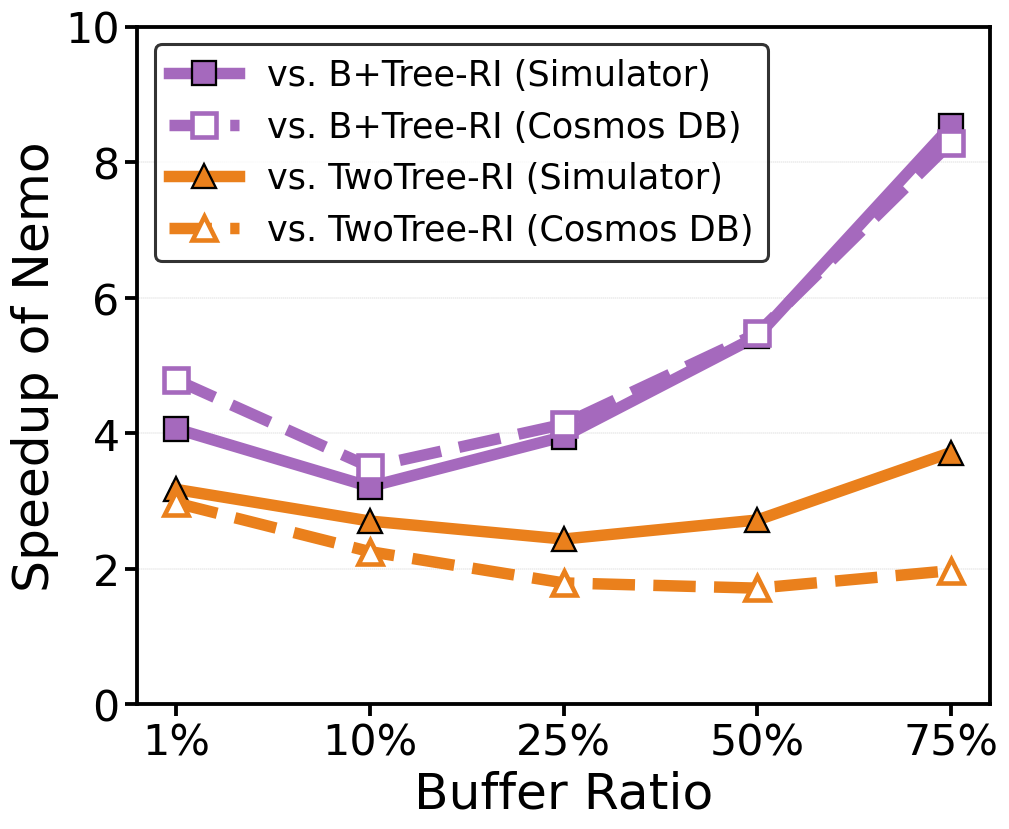}
     \caption{Speedup: Simulator vs.\ Cosmos DB}
    \label{fig:sim_store_b}
  \end{subfigure}
  \vspace{-0.08in}
  \caption{Performance on a Simulated Remote Store}
  \label{fig:sim_store}
  \Description{Left: grouped throughput bars on a log scale over buffer ratios from 1 to 100 percent for B+Tree-RI, TwoTree-RI, and Nemo on a simulated store without request-rate limits; Nemo leads at every ratio, and its margin over B+Tree-RI exceeds eight times at large buffers. Right: lines of Nemo's speedup over each baseline on the simulator and on Cosmos DB; for B+Tree-RI the two curves nearly coincide, differing only at the 1 percent buffer.}
\end{figure}

This experiment separates \proposed's architectural benefits from storage-side rate limiting. Cosmos DB's request-rate cap penalizes systems that issue more expensive requests, so we repeat the buffer-size experiment on a \textit{\textbf{simulated store}} that serves every request with a stable latency and imposes no cap. \figref{fig:sim_store_a} shows the same picture as \figref{fig:exp1_f1}: \proposed leads at every buffer ratio, its margin over \btreeretain grows with the buffer to well over 8$\times$, and at 100\% \twotreeretain catches up while \btreeretain still lags because its data and index pages do not fit. \proposed's speedup therefore comes from issuing fewer and smaller remote requests, not from throttling. \looseness=-1

\figref{fig:sim_store_b} plots \proposed's speedup over each baseline on the simulator and on Cosmos DB. Over \btreeretain the two curves nearly coincide; they differ only at the 1\% buffer, where \btreeretain issues the most requests and is throttled most on Cosmos DB.

\subsubsection{Ablation: Buffer Replacement Policy} \label{sssec:evict_policy} 

\begin{figure}
  \vspace{-0.1in}
  \centering
  \begin{subfigure}{0.35\linewidth}
    \centering
    \includegraphics[width=\linewidth]{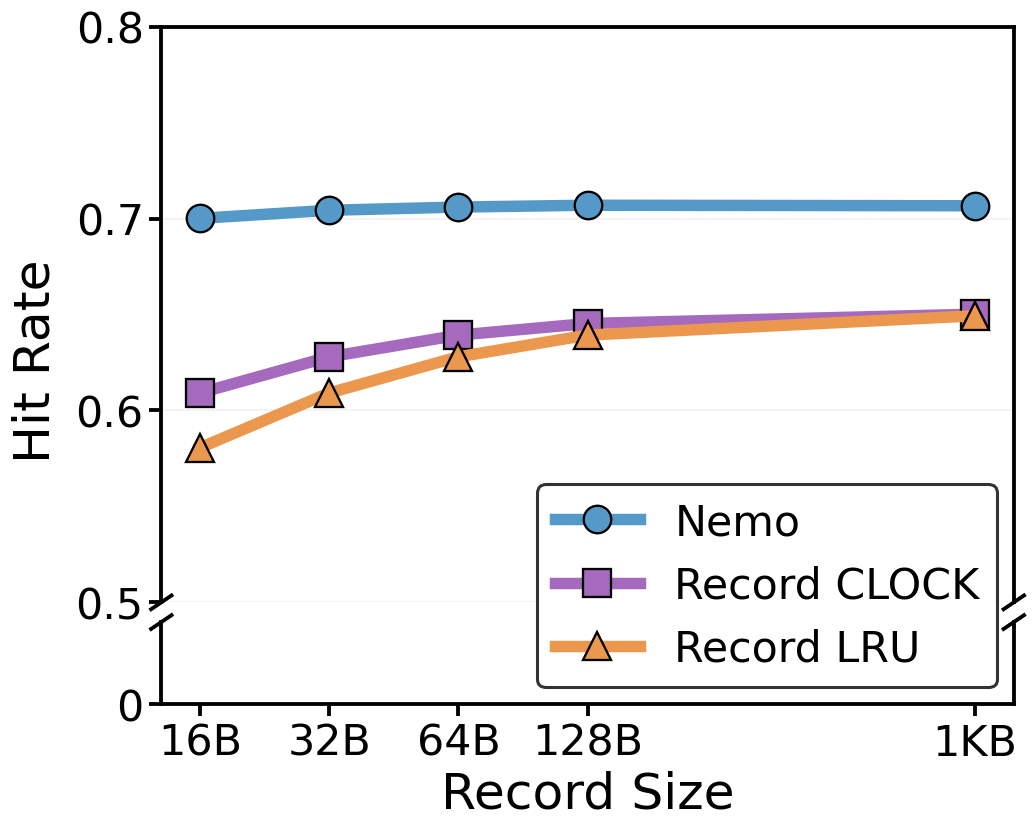}
     \caption{Hit Rate Over Different Methods}
    \label{fig:evict_policy_a}
  \end{subfigure}\quad%
  \begin{subfigure}{0.35\linewidth}
    \centering
    \includegraphics[width=\linewidth]{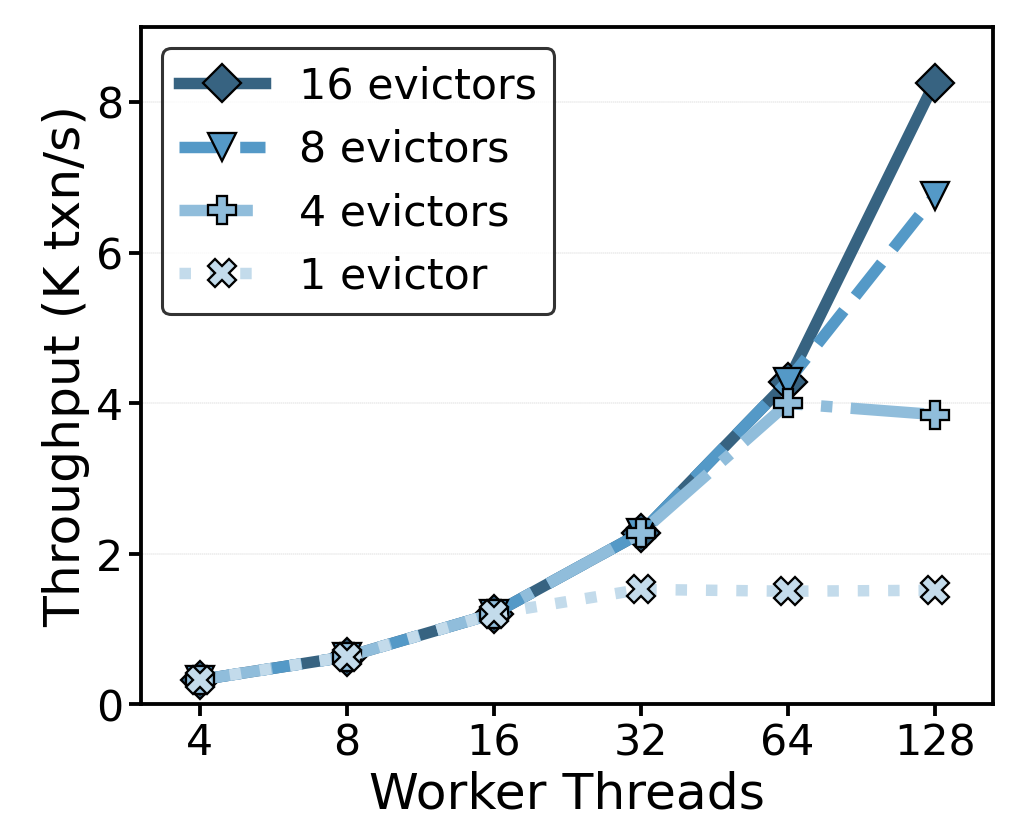}
     \caption{Nemo Eviction Scalability}
    \label{fig:evict_policy_b}
  \end{subfigure}
  \vspace{-0.08in}
  \caption{Buffer Replacement Policy}
  \label{fig:evict_policy}
  \Description{Left: hit rate versus record size from 16 bytes to 1 KB for Nemo, record-level CLOCK, and record-level LRU under the same memory budget; Nemo stays flat while CLOCK and LRU fall below it, most for small records. Right: throughput in thousands of transactions per second versus 4 to 128 worker threads, with 1, 4, 8, and 16 eviction threads; one eviction thread levels off near 1.5 thousand transactions per second from 32 workers, while 16 eviction threads scale linearly.}
\end{figure}

We compare \proposed's replacement policy (\secref{ssec:buf_replace}) with record-level CLOCK and LRU, on YCSB point lookups (10.24~M rows, $\theta=0.9$, 10\% buffer, 16 workers, read-only), charging each policy's per-record metadata (\proposed 5 bits, CLOCK 8 bytes, LRU 16 bytes) against the same buffer budget.

\boldtext{Space Efficiency} (\figref{fig:evict_policy_a}). \proposed's hit rate stays flat across record sizes, whereas CLOCK and LRU fall well below it, especially for small records: for a 16-byte record their pointers add 50--100\% to the footprint, so they hold only about 70\% and 50\% as many records under the same budget. \proposed thus matches or beats CLOCK and LRU in hit rate while spending the least memory per record. \looseness=-1

\boldtext{Eviction Scalability} (\figref{fig:evict_policy_b}). We fix 128-byte records and raise the number of workers, and hence eviction demand, to stress \proposed's parallel eviction. With one eviction thread, throughput flattens at around 1.5K txn/sec from 32 workers, as workers wait for free slots; more eviction threads raise the ceiling proportionally, and 16 threads scale linearly. Eviction parallelizes because each task touches only one root-to-leaf path (\secref{ssec:buf_replace}), so tasks on different leaves run concurrently.

\subsection{Negative Search}
\label{ssec:negative_search}

We study point and range queries at a 30\% buffer ratio, adding a hash-indexed record cache baseline.

\subsubsection{Negative Point Search} \label{sssec:n_point}

\begin{figure}[t]
  \vspace{-0.1in}
  \centering
  \begin{subfigure}{0.35\linewidth}
    \includegraphics[width=\linewidth]{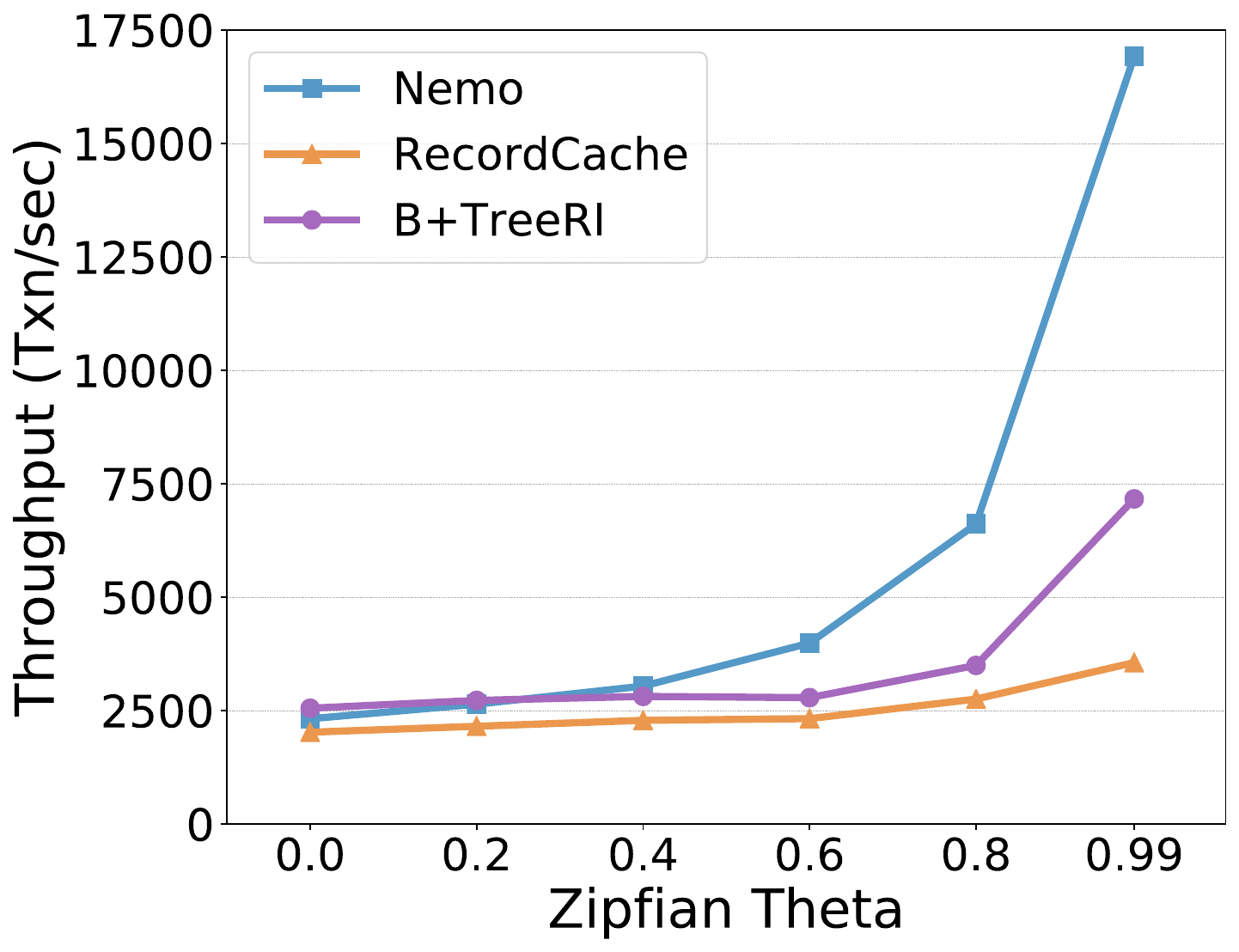}
    \caption{Vary Skewness (Point)}
    \label{fig:neg_point_vary_skew}
  \end{subfigure}\quad%
  \begin{subfigure}{0.35\linewidth}
    \includegraphics[width=\linewidth]{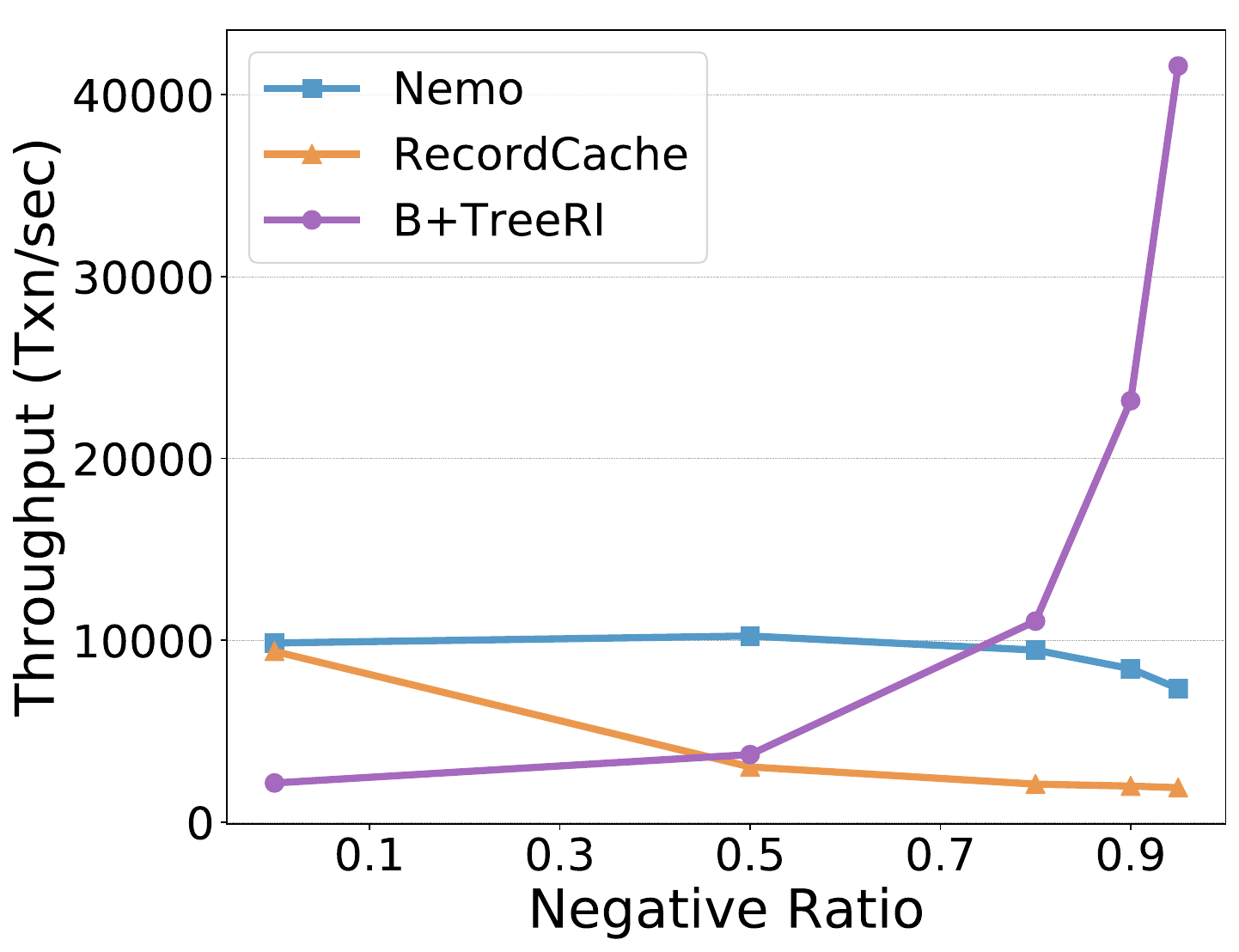}
    \caption{Vary Negative Ratio (Point)}
    \label{fig:neg_point_vary_neg_ratio}
  \end{subfigure}

  \begin{subfigure}{0.35\linewidth}
    \includegraphics[width=\linewidth]{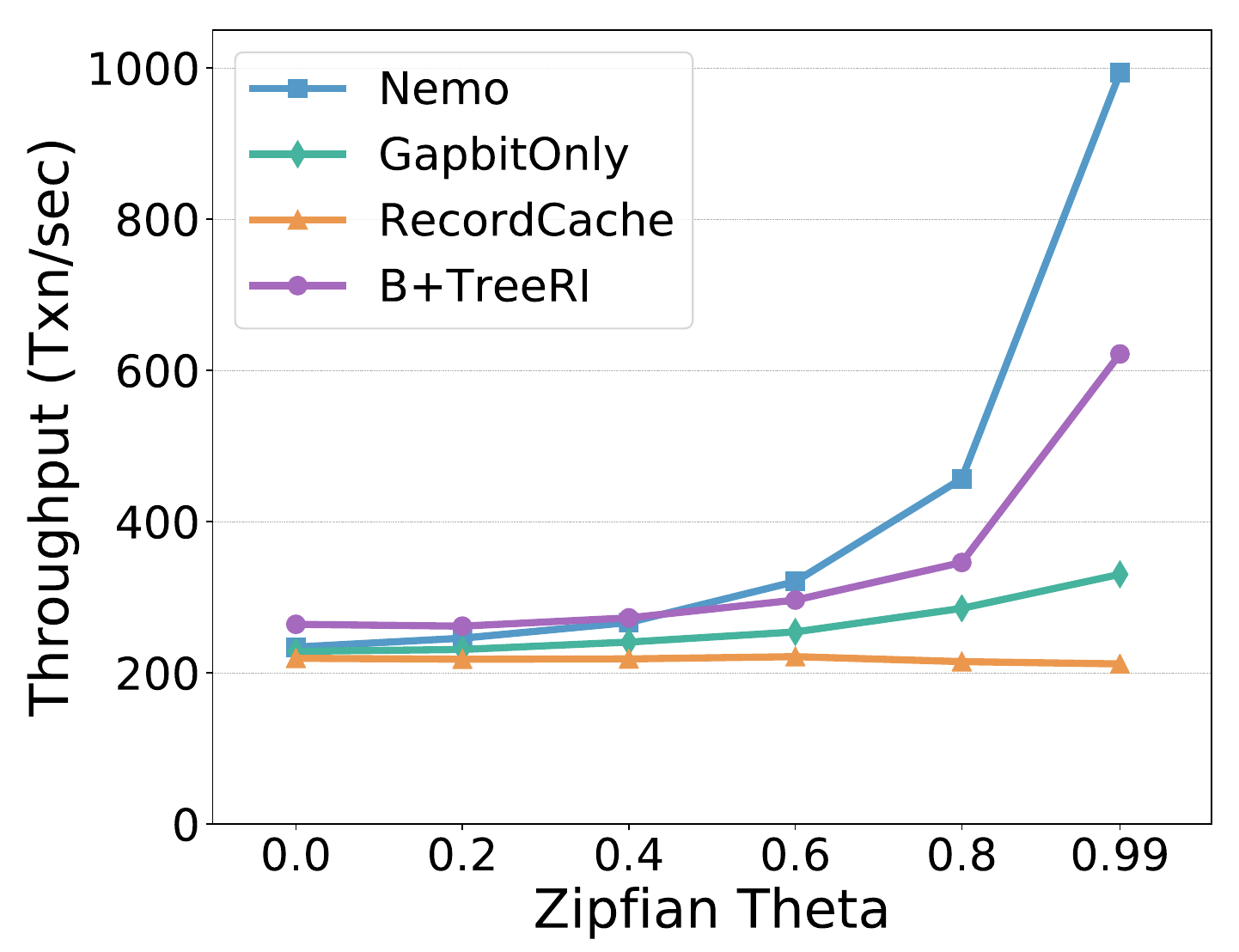}
    \caption{Vary Skewness (Range)}
    \label{fig:neg_range_vary_skew}
  \end{subfigure}\quad%
  \begin{subfigure}{0.35\linewidth}
    \includegraphics[width=\linewidth]{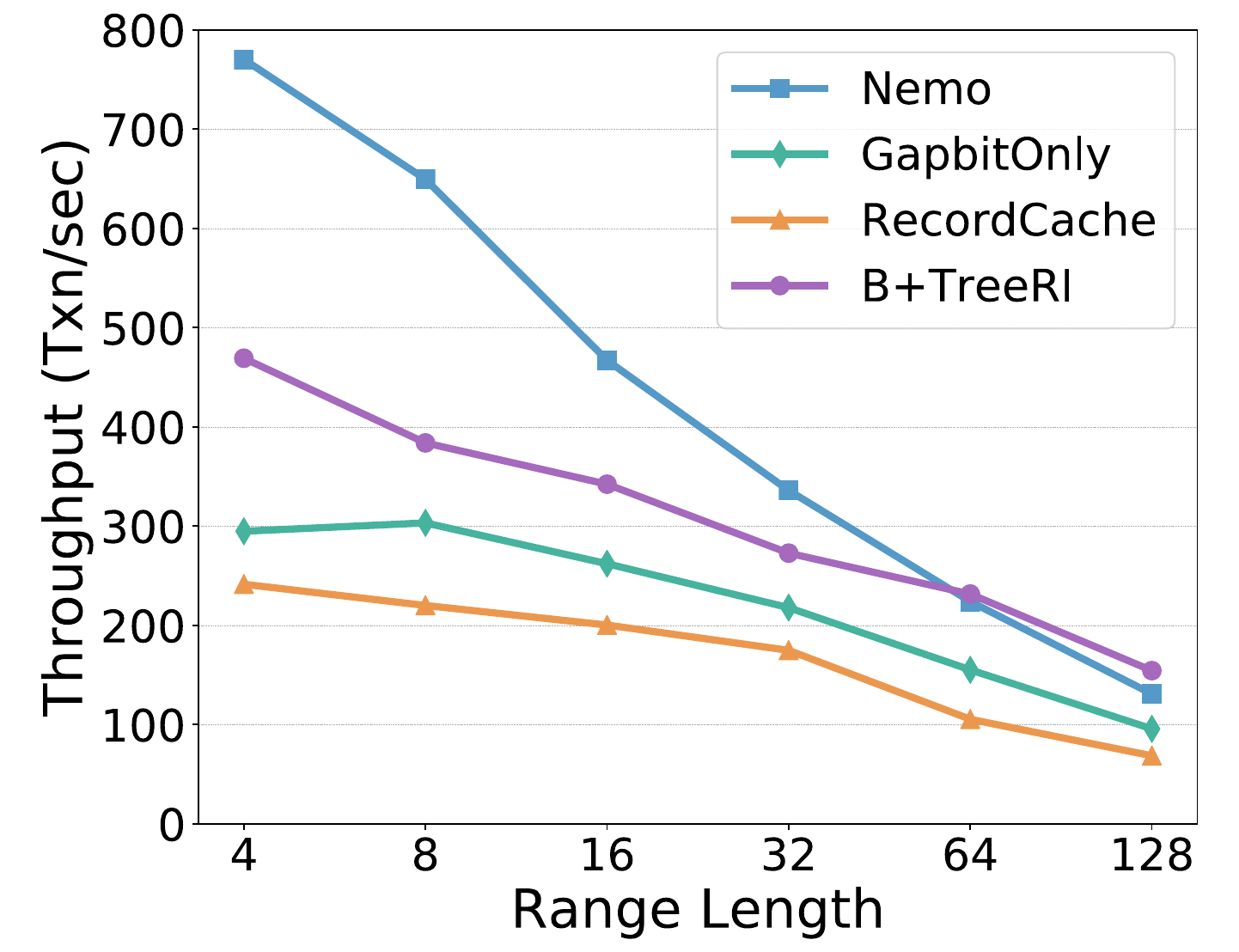}
    \caption{Vary Range Size (Range)}
    \label{fig:neg_range_vary_range_size}
 \end{subfigure}
  \vspace{-0.08in}
  \caption{Negative Search (Buffer Ratio 30\%)}
  \Description{Four line charts of throughput. Charts (a) and (c) vary Zipfian theta from 0 to 0.99, chart (b) varies the share of negative queries from 0 to 100 percent, and chart (d) varies the range length from 4 to 128. Point-search charts compare Nemo, a record cache, and B+Tree-RI; range-search charts add a GapbitOnly variant. Nemo's lead over the record cache grows with skew, up to 4.76 times for point queries; B+Tree-RI overtakes Nemo when most point queries are negative or ranges are very long, and throughput falls for every design as ranges get longer.}
\end{figure}

We fix the negative-query ratio ($\frac{\textit{\# Negative Queries}}{\textit{\# Total Queries}}$) at 50\% and vary query skew. In \figref{fig:neg_point_vary_skew}, \proposed consistently outperforms record cache: speedup rises from 1.15$\times$ to 4.76$\times$ as skew increases. Record cache only caches positive records, so all negative queries miss and incur remote probes. Although \btreeretain resolves negative queries locally by caching index pages, \proposed outperforms \btreeretain by up to 2.4$\times$ because it caches at record granularity, whereas \btreeretain caches at page granularity, coupling hot and cold data.

We then vary the negative ratio (0--100\%) under skew ($\theta=0.9$). In \figref{fig:neg_point_vary_neg_ratio}, \proposed consistently outperforms record cache by up to 4.5$\times$ for the same reason. When the negative ratio is 0, record cache matches \proposed since all queries are positive and both cache hot data at record granularity. \proposed outperforms \btreeretain when the negative ratio is low to moderate because of its higher cache efficiency. When the negative ratio is very high, \btreeretain significantly outperforms \proposed because it encodes negative information in cached index pages with no extra space overhead, whereas \proposed must cache negatives explicitly as tombstones.

\subsubsection{Range Search} \label{sssec:n_range}
Range queries are more complex than point queries because many are partially negative, containing both positive records and negative gaps. For example, a range over an integer key in [100, 200) that returns 30 records internally accesses 30 positive and 70 negative records. We set the key domain to twice the number of records (domain size = 2 million, data size = 1 million), so each range accesses negative records with 50\% probability. We add a \textit{GapbitOnly} baseline: a \proposed variant that encodes negative information with only gap bits, not tombstones.

We fix the range length at 8 (each query returns \(\approx\)4 records) and vary query skew. \figref{fig:neg_range_vary_skew} shows that under high skew \proposed outperforms record cache, \btreeretain, and \textit{GapbitOnly} by up to 4.70$\times$, 1.60$\times$, and 3.01$\times$. \proposed surpasses \btreeretain for the same reason as in negative point search. Record cache sustains low throughput because it cannot serve range queries from cache. \textit{GapbitOnly} gains slightly from gap bits but still suffers low cache efficiency: without cached negative boundaries, it needs remote probes for such ranges.

We then vary the range length from 4 to 128 under skew (\(\theta=0.9\)). In \figref{fig:neg_range_vary_range_size}, \proposed consistently outperforms record cache and \textit{GapbitOnly} by caching negative information more effectively. \proposed also outperforms \btreeretain for small to moderate ranges thanks to finer caching granularity. For large ranges (e.g., 128), \btreeretain surpasses \proposed because it also caches negative information and each query now spans multiple pages, reducing \proposed's fine-granularity advantage. Throughput drops for all methods as range length increases, since larger ranges raise miss probability and remote requests transfer more data.

Under random queries or very high negative ratios, tombstones can displace regular records from the buffer. Our experiments expose the resulting crossover with \btreeretain, whose resident index pages encode absence without extra cache space, whereas \proposed caches negative results explicitly. For point queries, \btreeretain is about 10\% faster under uniform access (\figref{fig:neg_point_vary_skew}), but \proposed overtakes it as skew increases. Similarly, \proposed leads as the negative ratio increases until roughly 75\%, after which \btreeretain becomes decisively faster (\figref{fig:neg_point_vary_neg_ratio}). Range queries show the same crossover ($\theta \approx 0.4$; \figref{fig:neg_range_vary_skew}). Thus, \btreeretain is preferable when negative results have little reuse or overwhelmingly dominate the workload. Otherwise, \proposed's record-granularity advantage outweighs the tombstone overhead. This overhead is bounded: each tombstone carries only 5 metadata bits and no payload, while the second-chance policy evicts cold tombstones to limit cache pollution. \looseness=-1

\subsection{Phantom Protection} \label{ssec:phantom_protection}

We compare \proposed against \pcl~\cite{lomet2009locking}. Both methods use the same indexing and caching scheme. The workload follows YCSB-E~\cite{cooper2010benchmarking} with concurrent range scans and inserts under skew ($\theta=0.99$) and a 30\% buffer ratio. We disable logging to isolate contention and prioritize read over write locks, keeping scan throughput similar across methods so insert throughput reflects conflict intensity.

\begin{figure}[t]
  \vspace{-0.1in}
  \centering
  \begin{subfigure}{0.35\linewidth}
    \includegraphics[width=\linewidth]{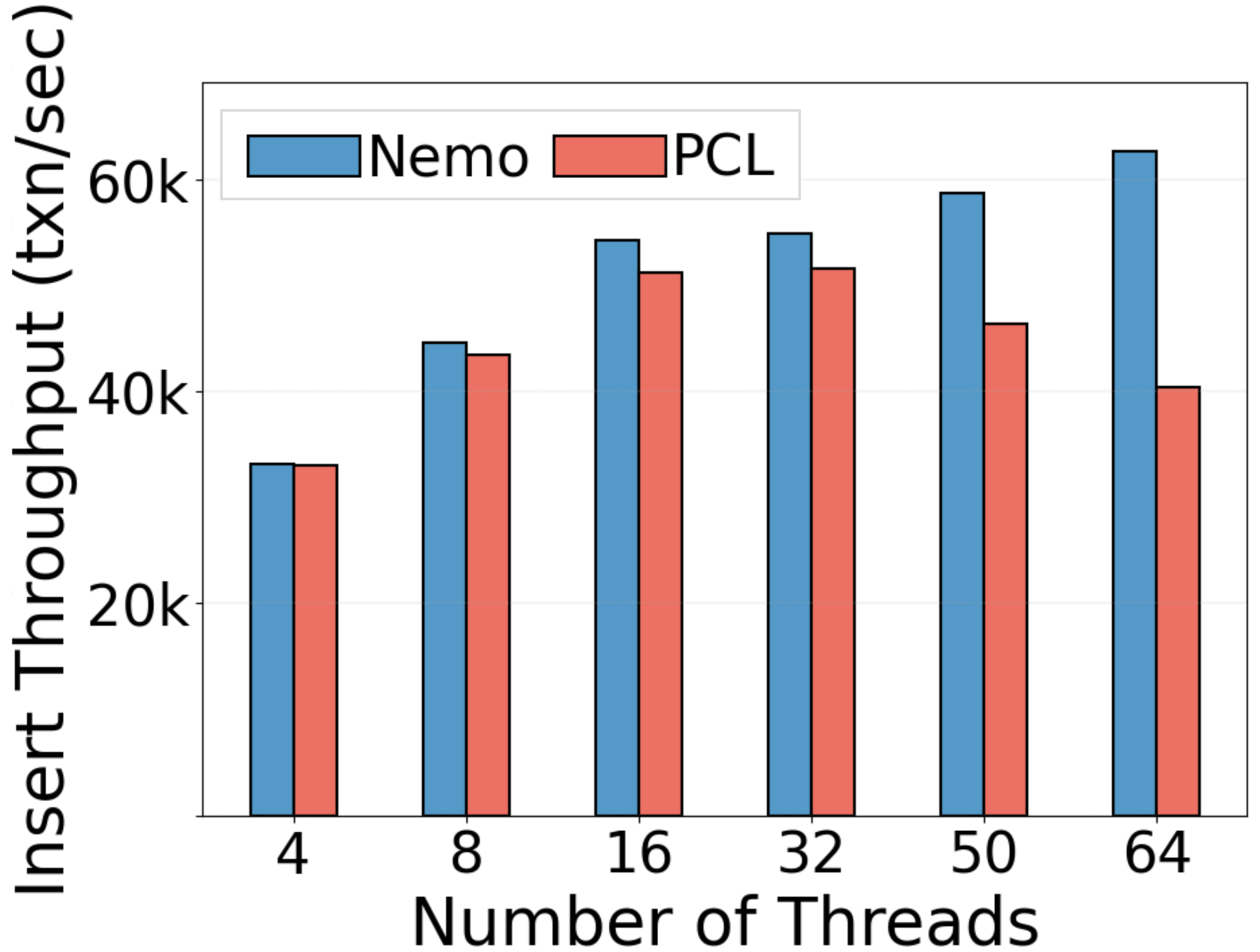}
    \caption{Insert Throughput}
    \label{fig:pp_scalability_a}
  \end{subfigure}\quad%
  \begin{subfigure}{0.35\linewidth}
    \includegraphics[width=\linewidth]{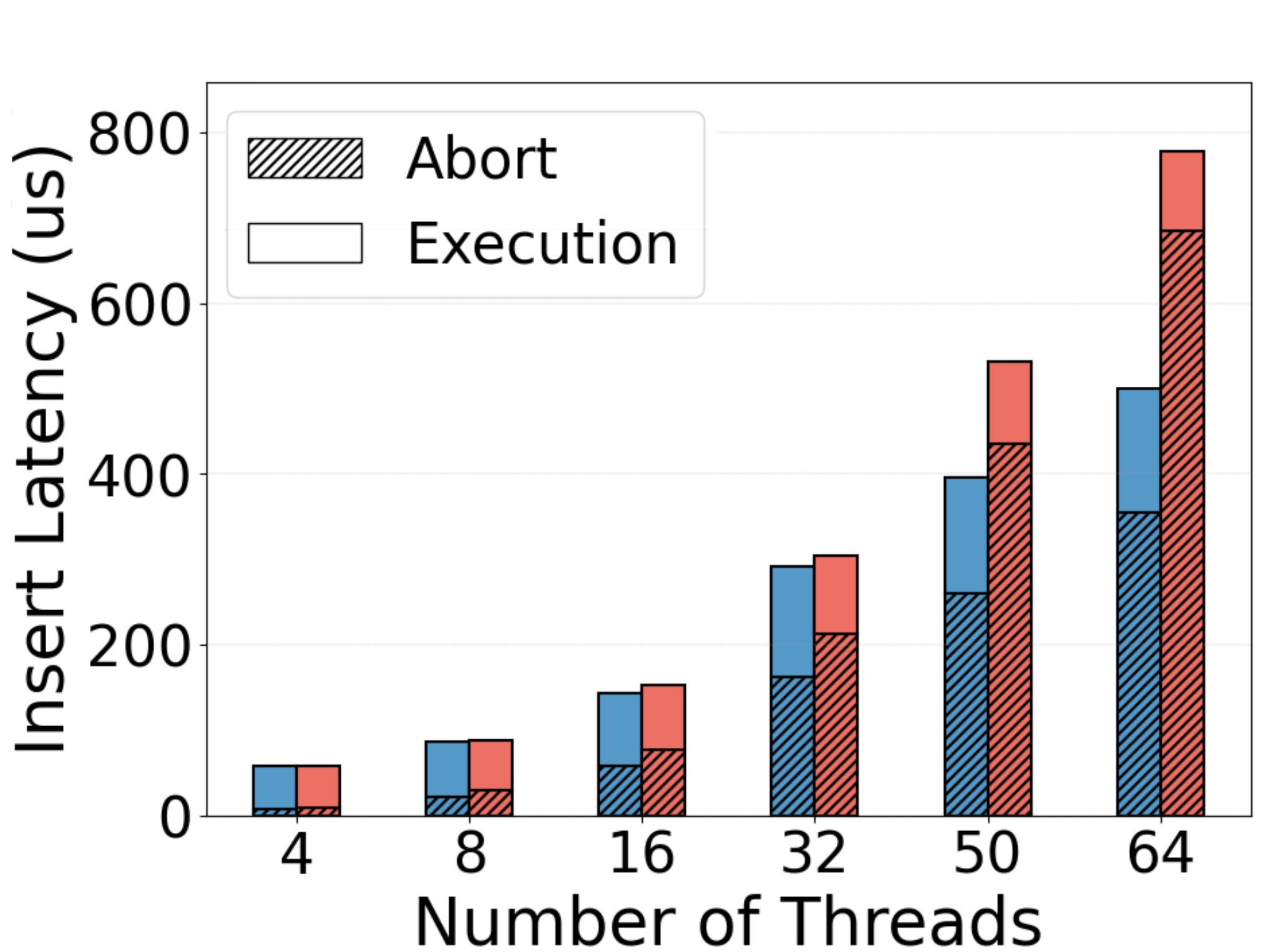}
    \caption{Insert Average Latency}
    \label{fig:pp_scalability_b}
  \end{subfigure}
  \vspace{-0.08in}
  \caption{Phantom Protection (Varying Concurrency)}
  \label{fig:pp_scalability}
  \Description{Left: insert throughput of Nemo and PCL for 4 to 64 threads; both scale similarly up to 32 threads, after which Nemo reaches 63 thousand transactions per second at 64 threads while PCL declines to about 40 thousand. Right: average insert latency split into abort and execution time; PCL's abort time grows sharply with more threads, reaching 685 microseconds at 64 threads.}
\end{figure}

\boldtext{Concurrency.} We use equal numbers of scan and insert threads, varying the total thread count from 4 to 64. \pcl uses 100~KB partitions. In \figref{fig:pp_scalability_a}, at low concurrency, both methods scale similarly because lock conflicts are rare. Beyond 32 workers, the two methods diverge: \proposed scales to 63K txn/sec at 64 workers, whereas \pcl's throughput peaks and declines to 40K (1.55$\times$ lower). \pcl's coarser partitions generate more false conflicts, while \proposed confines conflicts to the queried range. Exponential-backoff retries cost far more than successful inserts; \pcl's higher abort rate raises per-insert abort time to 685~\us at 64 workers (\figref{fig:pp_scalability_b}).

\begin{figure}[t]
  \vspace{-0.1in}
  \centering
  \begin{subfigure}{0.35\linewidth}
    \includegraphics[width=\linewidth]{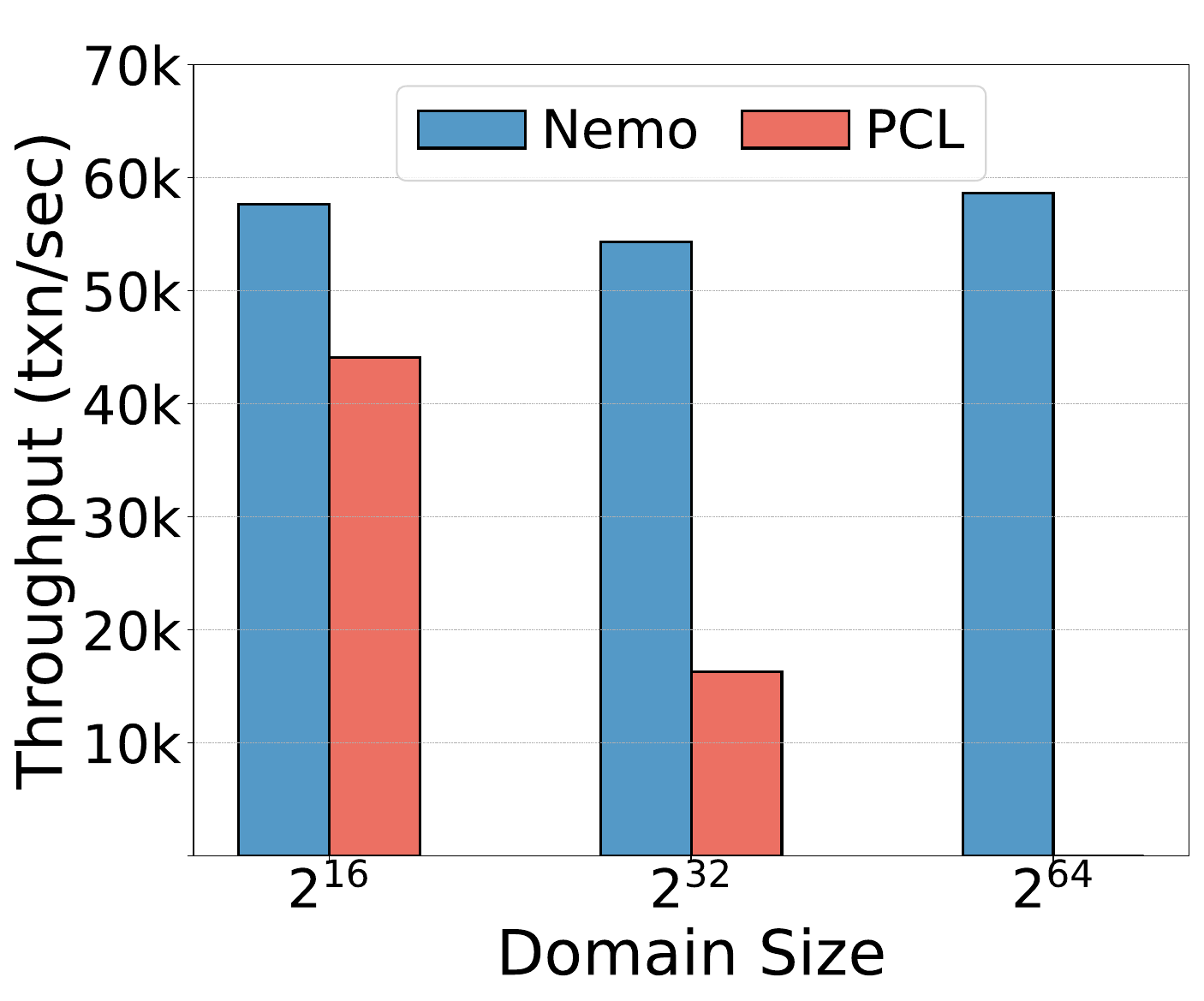}
    \caption{Insert Throughput}
    \label{fig:pp_f1_a}
  \end{subfigure}\quad%
  \begin{subfigure}{0.35\linewidth}
    \includegraphics[width=\linewidth]{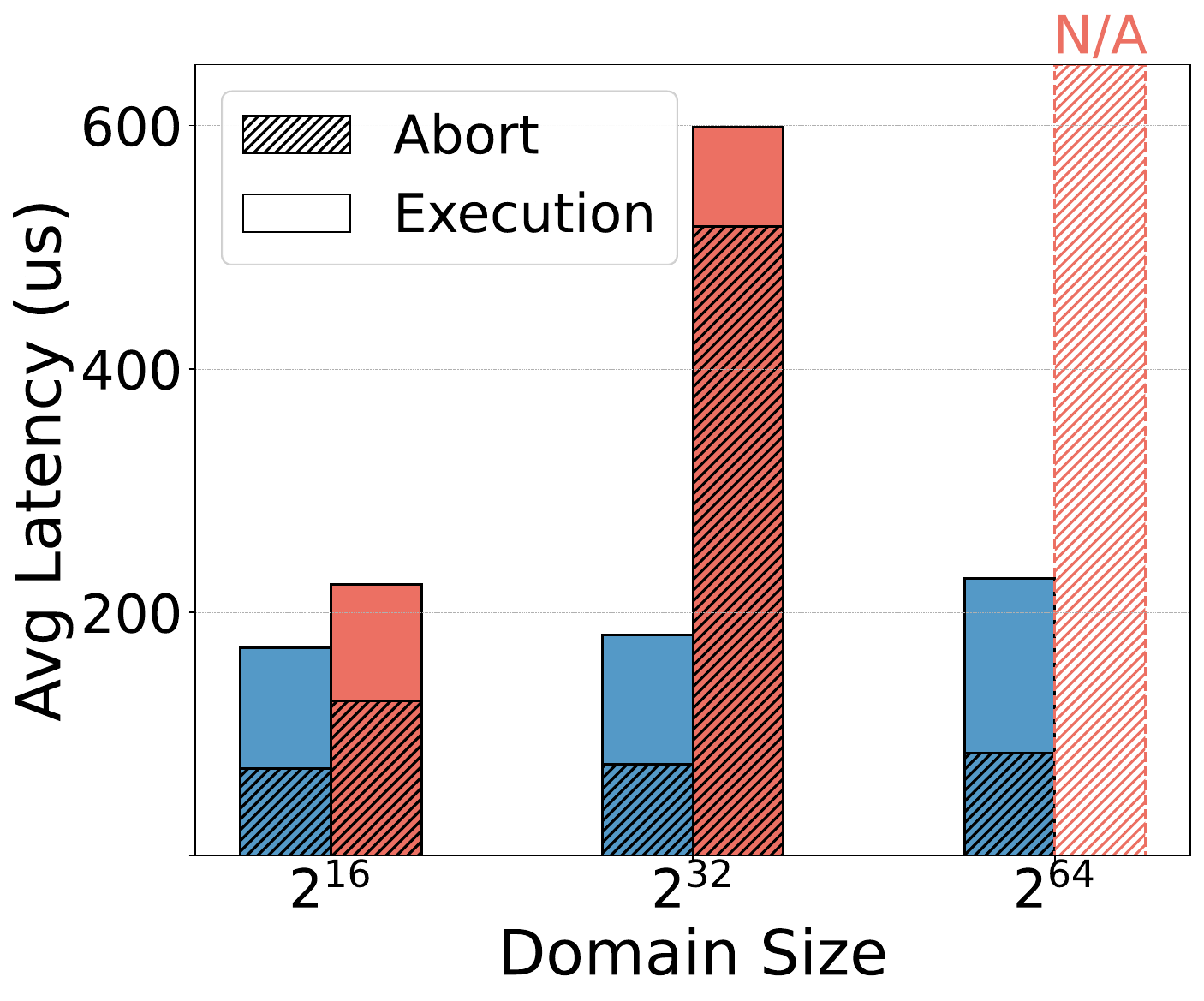}
    \caption{Insert Average Latency}
    \label{fig:pp_f1_b}
  \end{subfigure}
  \vspace{-0.08in}
  \caption{Phantom Protection (Varying Domain Size)}
  \label{fig:pp_domain}
  \Description{Left: insert throughput of Nemo and PCL for key domains of 2 to the 16th, 2 to the 32nd, and 2 to the 64th; Nemo stays nearly constant while PCL drops as the domain grows and has no result at 2 to the 64th, where one partition lock blocks all inserts. Right: average insert latency split into abort and execution time; PCL's abort latency rises with the domain size, and its 2 to the 64th bar is marked N/A.}
\end{figure}

\boldtext{Domain Size.} We study sensitivity to key-domain size, which drives partition granularity. We fix the dataset at roughly 1~GB and the thread count at 32. We vary the key domain to $2^{16}$, $2^{32}$, and $2^{64}$ (short, int, and long integers). The lock table capacity is 30\% of the buffer size. Since \pcl is sensitive to partition granularity, we tune it by selecting the smallest partition size whose locks fit in the lock table. In \figref{fig:pp_f1_a}, \proposed consistently outperforms \pcl, with the speedup increasing as the domain grows. For \pcl, larger domains force the use of a larger partition size to meet the lock-table space constraint. Thus, partitions cover more unrelated records, increasing contention (consistent with the rising abort latency in \figref{fig:pp_f1_b}). At $2^{64}$, \pcl requires partitions exceeding 2.1 trillion keys; a single partition lock thus covers nearly all hot data, blocking all inserts. In contrast, \proposed precisely locks the predicate range (average length 8), orders of magnitude smaller than \pcl's partitions and insensitive to the domain size.

\section{Related Work}
\label{sec:related}

\boldtext{Disk-Oriented Index.} Conventional disk-oriented indexes like B+trees~\cite{ubiquitousBtree, postgres} operate at page granularity. While recent work optimizes page-structured indexing~\cite{Levandoski2013bwtree}, these designs suffer low cache efficiency under skew and read/write amplification. Several works~\cite{zhou2023two-tree, anti-caching, Siberia, Treeline, ONeil1996LSM, rocksdb2021} improve memory utilization. Siberia~\cite{Siberia} migrates data between memory-optimized and page-based stores in Hekaton~\cite{Diaconu2013HekatonSS}. However, its page organization causes network amplification, and its offline migration adapts poorly to workload shifts. It also lacks phantom protection. \twotree~\cite{zhou2023two-tree} indexes hot and cold data in separate trees with record migration. Yet, it incurs network amplification by loading index and data pages. Anti-Caching~\cite{anti-caching} keeps hot records in memory and evicts cold data, but requires in-memory secondary indexes and metadata for all evicted records.

\boldtext{Indexes in Cloud Databases.} Modern databases~\cite{verbitski2017amazon, zhou2021foundationdb, Levandoski2011DeuteronomyTS, socrates, PolarDB, corbett12spanner} adopt storage disaggregation. Many systems, such as Aurora~\cite{verbitski2017amazon} and Socrates~\cite{socrates}, directly follow the page-based index design, replacing disks with a replicated storage service. Others~\cite{zhou2021foundationdb, Levandoski2011DeuteronomyTS} split the storage engine into a compute-layer transactional component and a storage-side data component. Deuteronomy~\cite{Levandoski2011DeuteronomyTS} decouples the two layers via record-based APIs and hides storage-side indexing (e.g., Bw-tree~\cite{Levandoski2013bwtree}) from the compute layer. However, its hash-indexed record cache cannot accelerate range queries. Moreover, its partition-covering locks~\cite{lomet2009locking} provide storage-agnostic phantom protection but amplify lock ranges and are sensitive to partitioning and skew.

\boldtext{Caching over Fast Remote Memory.}
DEX~\cite{lu2024dex} is an RDMA-based index for disaggregated memory. Both DEX and \proposed keep a hot tree in compute memory, but DEX caches part of the persistent global index at page granularity, whereas \proposed builds an ephemeral tree over only hot records and never loads the global index locally. AIFM~\cite{ruan2020aifm} manages far memory at object granularity. It merely swaps application heap objects under memory pressure and targets general-purpose memory tiering, whereas \proposed must support transactional caching and write buffering after index pushdown. Likewise, an in-memory key-value store (e.g., Redis~\cite{redis}, Memcached~\cite{memcached}) can hold cached records but is not itself an OLTP buffer for larger-than-memory datasets: supporting range completeness, negative search, phantom protection and durable write buffering would require adding these database semantics.

\section{Conclusion}
\label{sec:conclusion}

\proposed is a record buffer with index pushdown for storage-disaggregated OLTP. It avoids network amplification, low cache efficiency, and serial round-trips of page-based designs. It accelerates range queries using the cache and mitigates negative search degradation. It provides efficient phantom protection that requires no knowledge of the storage-side data layout. \proposed outperforms \btree by up to 11.42$\times$ on positive lookups and record caches by up to 4.76$\times$ on negative queries. For phantom protection, it stays stable and beats \pcl, which degrades sharply as key domains grow.

\begin{acks}
We thank Hokeun Cha and the anonymous reviewers for their constructive suggestions. This work was supported in part by NSF award IIS-2144588.
\end{acks}

\bibliographystyle{ACM-Reference-Format}
\bibliography{base,concurrency}

\end{document}